\documentclass[
twocolumn,
superscriptaddress,
 amsmath,amssymb,
 aps,
prb,
floatfix,
]{revtex4-2}

\usepackage[colorlinks]{hyperref}
\usepackage{amsmath}
\usepackage{adjustbox}
\usepackage{graphicx}
\usepackage{dcolumn}
\usepackage{bm}

\usepackage[svgnames]{xcolor}

\newcommand{\R}{\textbf{R}}
\newcommand{\cdag}{\hat{c}^{\dagger}}
\renewcommand{\c}{\hat{c}^{\phantom{\dagger}}}

\newcommand{\bairo}{Ba$_2$IrO$_4$}
\newcommand{\jeff}{$j_{\mathrm{eff}}=1/2$}
\newcommand{\JEFF}{$j_{\mathrm{eff}}=3/2$}

\newcommand\numberthis{\addtocounter{equation}{1}\tag{\theequation}}

\begin{document}

\preprint{APS}

\title{\texorpdfstring{The rich phase diagram of the prototypical iridate \bairo:\\ 
Effective low-energy models and metal-insulator transition}{Ba2IrO4}}

\author{Francesco Cassol}
 \thanks{These authors contributed equally.}
 \affiliation{Institut de Minéralogie, de Physique des Matériaux et de Cosmochimie, Sorbonne Université}
\author{Léo Gaspard}
 \thanks{These authors contributed equally.}
 \affiliation{Laboratoire de Chimie et Physique Quantiques, Université Toulouse III - Paul Sabatier}
\author{Michele Casula}
 \affiliation{Institut de Minéralogie, de Physique des Matériaux et de Cosmochimie, Sorbonne Université}
\author{Cyril Martins}
\thanks{Corresponding author}
\email{cyril.martins@irsamc.ups-tlse.fr}
 \affiliation{Laboratoire de Chimie et Physique Quantiques, Université Toulouse III - Paul Sabatier}
\author{Benjamin Lenz}
\thanks{Corresponding author}
 \email{benjamin.lenz@sorbonne-universite.fr}
\affiliation{Institut de Minéralogie, de Physique des Matériaux et de Cosmochimie, Sorbonne Université}

\date{\today}

\begin{abstract}
\begin{description}
\item[Abstract]
In the quest of new exotic phases of matter due to the interplay of various interactions, iridates hosting a spin-orbit entangled \jeff\ ground state have been in the spotlight in recent years.
Also in view of parallels with the low-energy physics of high-temperature superconducting cuprates, the validity of a single- or few-band picture in terms of the $j_{\mathrm{eff}}$ states is key.
However, in particular for its structurally simple member \bairo,\ such a systematic construction and subsequent analysis of minimal low-energy models are still missing.
Here we show by means of a combination of different \textit{ab initio} techniques with dynamical mean-field theory that a three-band model in terms of Ir-$j_{\mathrm{eff}}$ states fully retains the low-energy physics of the system as compared to a full Ir-$5d$ model.
Providing a detailed study of the three-band model in terms of spin-orbit coupling, Hund's coupling and Coulomb interactions, we map out a rich phase diagram and identify a region of effective one-band metal-insulator transition relevant to \bairo.
Compared to available angle-resolved photoemission spectra, we find good agreement of salient aspects of the calculated spectral function and identify features which require the inclusion of non-local fluctuations.
In a broader context, we envisage the three- and five-band models developed in this study to be relevant for the study of doped \bairo\ and to clarify further the similarities and differences with cuprates.
\end{description}
\end{abstract}

\maketitle

\section{Introduction}
\begin{figure}[tb!]
\begin{minipage}{.47\textwidth}
\centering
\includegraphics[width=\textwidth]{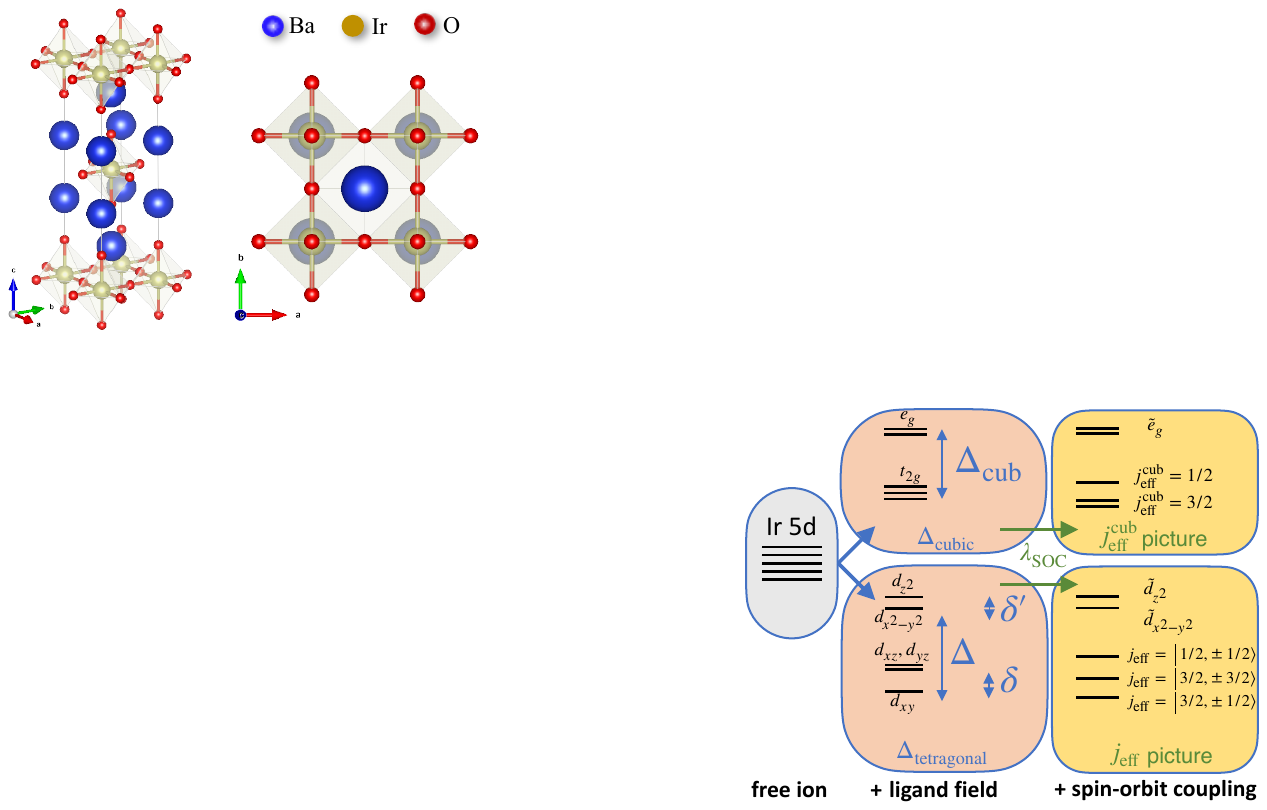}
\caption{
    K$_2$NiF$_4$-type crystal structure of \bairo. 
    The IrO$_2$ planes are responsible for its quasi-2D low-energy physics and in contrast to Sr$_2$IrO$_4$, the elongated IrO$_6$ octahedra are not rotated around the $c$-axis.
    }
\label{fig:Ba2IrO4_structure}
\end{minipage}
\end{figure}

Over the last decade, iridates have attracted a lot of interest, mainly due to their spin-orbit entangled $j_{\mathrm{eff}}$ ground state \cite{Kim2,Basal_afm_exp}.
It emerges as a consequence of the interplay of strong spin-orbit coupling (SOC), electronic Coulomb interactions and crystal field splitting and gives rise to exotic phases of matter such as quantum spin liquids, topological semimetals and spin-orbit entangled insulators \cite{Pesin_phase_diagram_nature,Topological_semimetal,Quantul_Liquid_theory,Spin_Liquid_Na4Ir4O4,Electronic_structure_of_iridates,Kitaev_begin,Kitaev_diagram,Wang2011,Anisotropy_exchange_Sr2IrO4,SOC_tuned_GS,Square_lattice_iridates}.\\
A particular focus has been set on Sr$_2$IrO$_4$, whose half-filled $j_{\mathrm{eff}}=1/2$ band suggests a description in terms of a twisted one-band Hubbard model \cite{Wang2011} with parallels to isostructural high-temperature superconducting cuprates.
However, despite similarities in their magnetic \cite{Okabe2011} and spectroscopic \cite{delaTorre2015,Arpes_Sr2IrO4,peng_electronic_2022} properties, no superconductivity has been reported to date for doped Sr$_2$IrO$_4$.
Possible explanations for this absence are twofold. 
First, rotations of the IrO$_6$ octahedra lead to a spin-canting of the Ir-momenta \cite{Crawford1994}, which results in a Dzyaloshinsky-Moriya term in the effective pseudospin model \cite{Kitaev_begin,Perkins2014}, absent in spin models of cuprates \cite{Coldea2001}.
Secondly, the validity of a simple single-band picture has been cast into doubt, questioning either the \textit{local} \cite{Arpes_Sr2IrO4, magnetization_sr2iro4_dxy} or the single-band nature \cite{Pavarini_Sr2IrO4_multiorbital} of the $j_{\mathrm{eff}}=1/2$ state upon doping.\\
Instead, Ba$_2$IrO$_4$, a closely related material with the same nominal $5d^5$ configuration on the iridium site, crystallizes in a K$_2$NiF$_4$-type crystal structure (space group I4/mmm)\cite{Okabe2011} like La$_2$CuO$_4$, see \autoref{fig:Ba2IrO4_structure}.
The combination of a tetragonal ligand-field splitting and strong spin-orbit coupling leads to the much-invoked $j_{\mathrm{eff}}$ picture inherited from Sr$_2$IrO$_4$.
In this picture, the $5d^5$ configuration realised in \bairo\ amounts to a single electron in the \jeff\ state, giving rise to a Mott insulating state when accounting for the strong electron-electron interactions within this half-filled band.
The absence of rotations of the IrO$_6$ octahedra leads to an in-plane antiferromagnet below a N\'eel temperature of $T_N\approx240$ K \cite{Okabe2011}.
This magnetic order is well described by a pseudospin model dominated by Heisenberg terms \cite{Katukuri2014,Hou2016} and thereby even closer to the situation realised in cuprates.\\
Both the magnetically ordered and the paramagnetic phases of \bairo\ are insulating \cite{Okabe2011,Optical_conducivity_Ba2IrO4,Moser,Uchida_ARPES}, the latter up to at least $300$ K \cite{Moser,Optical_conducivity_Ba2IrO4}. 
Angle-resolved photoemission studies have shed light on the band structure of both phases \cite{Moser, Uchida_ARPES}, but differed in their findings concerning changes in spectral features at the transition temperature.
Given the difference in the type of samples used - thin films \cite{Uchida_ARPES} and high-pressure synthesized bulk samples \cite{Moser} - spectral signatures of antiferromagnetic order in \bairo\ still need to be firmly established.
First theoretical studies including extensions of density functional theory (DFT+U) \cite{Moser,Uchida_ARPES} and its combination with dynamical mean-field theory (DFT+DMFT) \cite{Arita,Arita2} found good qualitative agreement with the most salient 
aspects
of ARPES spectra, but focused mainly on its antiferromagnetic phase.
These studies also suggested a half-filled $j_{\mathrm{eff}}=1/2$ ground state for Ba$_2$IrO$_4$, without rigorously justifying the range of validity of this picture.\\
In order to assess the supposed parallels to the low-energy physics realized in cuprates, the derivation of such minimal effective models for \bairo\ is crucial. 
Whereas this question has been addressed in case of the cuprates mainly with respect to the role of oxygen atoms \cite{ZSA,ZR1,Andersen}, it needs to be asked here first of all with respect to the impact of Ir-$e_g$ and \JEFF\ bands.

In this paper, we hence first study the electronic structure of \bairo\ in detail and derive different \textit{ab initio} models to describe its low-energy properties.
Following a top-down approach, we motivate the use of a three-band $j_{\mathrm{eff}}$ model by comparing its properties to the one of a complete five-band description in terms of total angular momentum $J$ states.
Then focusing on the former and restricting ourselves to paramagnetic phases, we investigate the metal-insulator transition as a function of interaction strength, spin-orbit coupling and temperature to establish a region of validity for a further reduction to a single-band \jeff\ model.
Using DMFT, we map out a phase diagram of our model to study the proximity of \bairo\ to topological and Mott transitions.
By comparing the resulting spectral functions to available angle-resolved photoemission spectroscopy data \cite{Moser,Uchida_ARPES} we can finally identify signatures of short-range spatial correlations that require a scheme going beyond a \textit{local} self-energy.
The remainder of the paper is organised as follows. 
In \autoref{section:model_construction} we construct two effective low-energy models of \bairo\ from first principles, which we study by means of DMFT in \autoref{section:validation}.
The interplay of spin-orbit coupling and Coulomb interaction and their impact on the Mott transition is analyzed in \autoref{section:spin_orbit} before comparing ARPES data from literature to calculations done for a parametrization of our model relevant to \bairo\ in \autoref{section:comp_with_exp}.
Finally, \autoref{section:summary} summarizes the main conclusions of the paper and discusses the parallels to the low-energy physics in cuprate.
    
\section{Model construction from first principles}
\label{section:model_construction}

    In this Section we derive from first principles two low-energy models for \bairo: a highly accurate five-band Ir-$5d$ model and an effective three-band Ir-$t_{2g}$ one, which will then be solved by means of DMFT.
    The models are defined by the following generic Hamiltonian: 
    \begin{equation}
            \label{eq:model_tot}
            \hat{H} = \hat{H}_{\text{0}} + \hat{H}_{\text{SOC}}+\hat{H}_{\text{int}},
    \end{equation}
    where the one-body terms $\hat{H}_{\text{0}}$ is defined by:
    \begin{equation}
            \label{eq:model_kin}
         \hat{H}_{\text{0}} = \sum_\sigma \sum_{i\R, j\R'} t^{\phantom{\dagger}}_{i\R, j\R'} \cdag_{\sigma i\R^{\phantom{\prime}}}\c_{\sigma j\R'} ,
    \end{equation}
    The indices $\R,\R'$ run over the Ir sites and  $\sigma$ denotes the electron spin. The labels $i, j$ refer to the five 5d orbitals on each site or only to the three $t_{2g}$ states, depending on the considered model. 
    $\hat{c}^{(\dagger)}_{\sigma j\R'}$ are electron (creation) annihilation operators and $t^{\phantom{\dagger}}_{i\R, j\R'}$ are the tight-binding  (TB) parameters. 
    Local on-site energies are included in the kinetic term (\autoref{eq:model_kin}) for $\R=\R'$.
    
    The SOC term $\hat{H}_{\text{SOC}}$ is purely local and defined using an isotropic spin-orbit coupling constant  $\lambda$ between the spin-orbitals $(i,\sigma)$ and $(j,\sigma')$ : 
          \begin{equation}
           \label{eq:model_SOC}
         \hat{H}_{\text{SOC}} =  \lambda \sum_{\sigma\sigma'}\sum_{ij\R}~\langle i\sigma|{\widehat{\mathbf{L}}}\cdot {\widehat{\mathbf{S}}}|j\sigma'\rangle~ \cdag_{\sigma i\R\phantom{'}}\c_{\sigma' j\R}.
        \end{equation}

    The Coulomb interaction $\hat{H}_{\mathrm{int}}$ is also local and defined in the extended Hubbard-Kanamori form \cite{Kanamori1,Correlations_and_J}:
    \begin{align}
            \label{eq:model_int}
         \hat{H}_{\text{int}} =  &\frac{1}{2} \sum_{\sigma} \sum_{ij} U_{ij} \hat{n}_{i\sigma}\hat{n}_{j\bar{\sigma}} + \frac{1}{2} \sum_{\sigma} \sum_{i \neq j} (U_{ij} - J_{ij}) \hat{n}_{i\sigma}\hat{n}_{j\sigma} \nonumber \\&
            - \frac{1}{2} \sum_{\sigma}\sum_{i\neq j} J_{ij} \left[ \cdag_{i\sigma}\c_{i\bar{\sigma}}\cdag_{j\bar{\sigma}}\c_{j\sigma} - \cdag_{i\sigma}\cdag_{i\bar{\sigma}}\c_{j\sigma}\c_{j\bar{\sigma}}\right],
    \end{align}
    where, for the sake of readability, we omitted the sum over $\R$.
    The first two terms are the density-density terms, representing the Coulomb repulsion between electrons with antiparallel and same spin respectively. 
    The last one includes the spin-flip and pair hopping terms. 
    
    All parameters have been calculated from first principles. 
    In particular, starting from a standard DFT reference calculation, we construct maximally localized Wannier functions (MLWFs) \cite{MLWF} and we estimate the spin-orbit coupling constant $\lambda$ through a fitting procedure \cite{Qiangqiang2023}. 
    From our localized basis set, we also evaluate the resulting Coulomb parameters via constrained random phase approximation (cRPA) \cite{cRPA}. 
   Computational details can be found in Appendix~\ref{appendix:Computational_details}.

  \subsection{Density functional theory band structure}
    \label{Sec:DFT}
    The DFT calculations, using the PBE functional \cite{PBE1996}, are performed starting from the experimental crystal structure of Ba$_2$IrO$_4$ in the primitive cell \cite{Isobe2012}. 
    The atomic positions within the cell are then relaxed up to a convergence factor of 10$^{-3}$ a.u. on the forces, which leads to a small reduction of the Ir-O bond length along the z direction (apical oxygen), enhancing the crystal field anisotropy, see Appendix~\ref{appendix:atom_coordinates}. 

    \begin{figure}[tb!]
    \centering
    \includegraphics[width=0.95\linewidth]{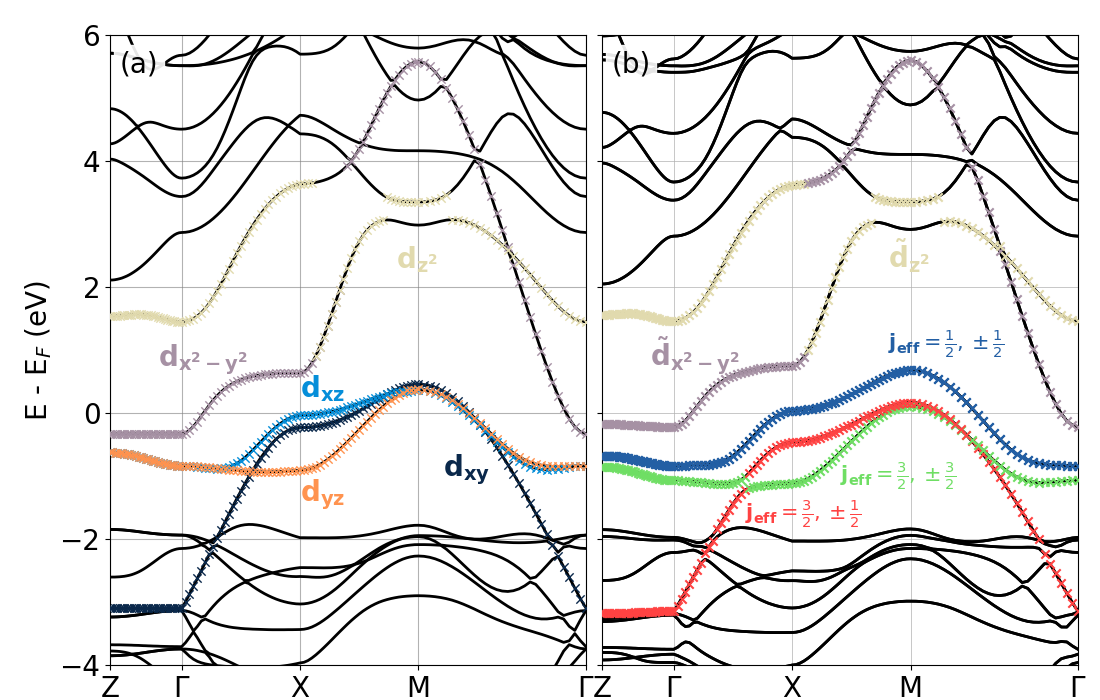}
    \caption{Kohn-Sham band structure of \bairo.
    DFT bands calculated without (a) and with (b) spin-orbit coupling using the PBE functional. The high-symmetry \textbf{k}-points chosen are $Z = (0, 0, \pi/c)$, $\Gamma = (0,0,0)$, $X = (\pi/a, 0, 0)$, and $M = (\pi/a, \pi/a, 0)$ in the Brillouin zone of the conventional unit-cell.  
    Colors represent the principal character of the bands with respect to the Ir 5d orbitals. 
        }
    \label{fig:ks_bandstructure_and_hopping}
    \end{figure}   
    
    The  Kohn-Sham band structure is shown in \autoref{fig:ks_bandstructure_and_hopping}. 
    In order to facilitate the comparison to photoemission spectra documented in literature, we refer here to the high-symmetry points in the conventional cell, i.e. X$=(\pi/a,0,0)$, M=$(\pi/a,\pi/a,0)$ and Z=$(0,0,\pi/c)$. 
    In the non-relativistic band structure shown in panel (a), the full $d$ manifold is clearly separated into $e_g$ and $t_{2g}$ manifolds. 
    Even if the $e_g$ bands are nearly empty, the $d_{x^2-y^2}$ band crosses the Fermi level. 
    A crystal-field anisotropy is clearly evident in the band structure, with the $d_{xy}$ band overall lower in energy and showing a wider dispersion. 
     Due to the layered structure, the Ir-5d bands are nearly dispersionless along the $\textbf{k}_z$ direction.

     \autoref{fig:ks_bandstructure_and_hopping} (b) reports the electronic structure including relativistic effects.
    The SOC splits the $t_{2g}$ states into two lower \JEFF\ bands and a \jeff\ band. 
    The effect of SOC on the $e_g$ manifold is rather small, but gives rise to the slightly more entangled states $\tilde{e}_g$. 
    Still, four bands cross the Fermi level, including the $\tilde{d}_{x^2-y^2}$, but, in contrast to DFT without SOC, the \jeff\ band is nearly half-filled $(n_{1/2}=1.23)$ and the \JEFF\ bands nearly completely filled $(n_{3/2,1/2}=1.83~; n_{3/2,3/2}=1.89)$. 
    Focusing only on the occupied $t_{2g}$ manifold, previous works \cite{Moser, Uchida_ARPES, Arita,Arita2} then considered an effective three-band model. 
    However, the choice of such a low-energy model for \bairo\ requires rigorously to confirm the weak impact of the $\tilde{e}_{g}$ manifold before integrating these degrees of freedom out. 
    In the following, we will then define a full five-band model and an Ir-$t_{2g}$ model and compare their properties at every step of the calculation. 
    
    \subsection{Parametrization of the models}
    \label{subsection:parametrization of the model}
    We now detail the parametrization of the two models built to describe the low-energy part of the electronic band structure of \bairo\ using MLWFs. 
    We stress that the Wannierization and subsequent cRPA calculations are different for the two models. 
    This is in particular visible in the spread of the $d_{xy}$ Wannier function, which is smaller in the five-band model than in the three-band model (4.21 \AA$^2$ vs. 4.41 \AA$^2$), whereas the spread of the other two $t_{2g}$ Wannier functions remains very similar (4.18 \AA$^2$ vs. 4.19 \AA$^2$). 
    This difference is explained by the presence of the $d_{x^2-y^2}$ Wannier function, which allows to disentangle its contribution to the $d_{xy}$ Wannier function and causes a smaller spatial extent of the latter.

\begin{figure}[tb!]
        \centering
            \includegraphics[width=0.95\linewidth]{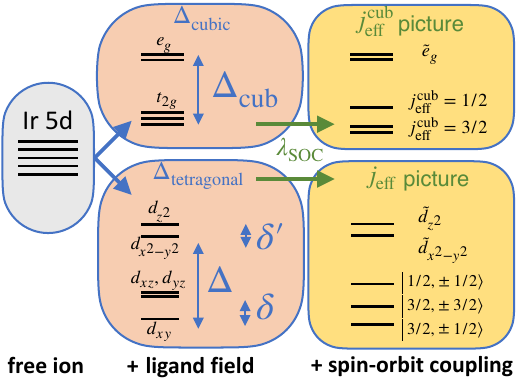}
            \caption{Energy level diagram for \bairo.
            Ligand-field splitting and spin-orbit coupling lead to the formation of spin-orbit entangled $j_{\mathrm{eff}}$ and $\tilde{e}_g$ states. 
            For $\Delta_{(\mathrm{cub})} \gg \lambda_{SOC}$, the SOC can be applied to the $t_{2g}$ manifold only, leading to the $j^{(\mathrm{cub})}_{\mathrm{eff}}$ states described in the text.  
            }
        \label{fig:LevelDiag}
    \end{figure}

    Despite the different Wannierizations, the TB models for both cases are very similar.
    In particular, the TB parameters within the $t_{2g}$ manifold differ at most by $10^{-2}$~eV since the lattice symmetry leads to a nearly absent $t_{2g}-e_g$ hybridization without SOC. 
    The hopping intensities decrease quickly as the distance increases. 
    The main hopping intensities are the nearest and next-nearest neighbor interactions.\\
    For the five-band model, the local spin-resolved Hamiltonian $H_{d}^{\text{loc}}$ splits into a block-diagonal form consisting of two 5$\times$5 blocks and can be parametrized in terms of only four parameters. 
    In their respective basis $\{ d_{xz} \pm, d_{yz} \pm, d_{xy} \mp, d_{z^2} \mp, d_{x^2-y^2} \mp \}$, where we denoted up (down) spins with $+$ ($-$), the 
    blocks read as: 
     \begin{equation}
        H_{d}^{\text{loc}} = \begin{pmatrix}
        \begin{array}{ccc|cc}
            \delta & \mp \frac{\lambda}{2} i & \frac{\lambda}{2}i & \pm\frac{\sqrt{3}\lambda}{2} & \mp\frac{\lambda}{2} \\[2pt]
            \pm\frac{\lambda}{2}i & \delta & \mp \frac{\lambda}{2} & -\frac{\sqrt{3}\lambda}{2}i & -\frac{\lambda}{2} i \\[2pt]
            -\frac{\lambda}{2}i & \mp\frac{\lambda}{2} & 0 & 0 & \mp\lambda i \\[2pt]
            \hline
            \rule{0pt}{1.25\normalbaselineskip}\pm\frac{\sqrt{3}\lambda}{2} & \frac{\sqrt{3}\lambda}{2}i & 0 & \Delta + \delta' & 0 \\[2pt]
            \mp\frac{\lambda}{2} & \frac{\lambda}{2}i & \pm\lambda i & 0 & \Delta
        \end{array}
        \end{pmatrix}
        \label{eq:somatrix}
    \end{equation}
    The three terms $\Delta$, $\delta$ and $\delta'$ are the so-called ligand field terms. 
    $\Delta$ accounts for the $t_{2g}-e_{g}$ splitting, $\delta$ ($\delta '$) encodes the splitting of the states within the $t_{2g}$ ($e_{g}$) sub-manifold, see \autoref{fig:LevelDiag}. 
    In general, $\delta$ and $\delta'$ define an effective tetragonal field. 
    The cubic (cub) case is recovered for $\delta=\delta'=0$ eV. 
    From our set of  Wannier functions, we obtain ligand field terms $\Delta = 3.14$ eV, $\delta = 0.24$ eV, and $\delta' = 0.2$ eV for the five-band model. 
    
    In the three-band model, the local Hamiltonian in \autoref{eq:somatrix} is restricted to the topmost $3\times3$ block $H_{t_{2g}}^{\text{loc}}$. 
    Thus, $\Delta$ and $\delta '$ are not present in the local Hamiltonian, and the computed value of $\delta=0.25$ eV is slightly enhanced  compared the full $5d$ one.
    We note that the sign of the ligand field component $\delta$  would correspond to the crystal field of a compressed octahedron, in contrast to the elongated IrO$_6$ octahedron found in \bairo. 
    This difference between crystal field and ligand field is due to the different environments of the apical and the in-plane oxygen atoms \cite{Figgis1999}.

    For the SOC constant, we find $\lambda=0.31$ eV in both models. Thus, $\lambda$ is an order of magnitude smaller than the crystal field splitting $\Delta$. As a result, comparing the full 5d model and the Ir-$t_{2g}$ one, we note that: 
    (i) The three eigenstates of $j_{\mathrm{eff}}$ character contain only small admixtures of the $d_{z^2}$ and $d_{x^2-y^2}$ orbitals. 
    (ii) The $\widetilde{e}_g$ states are nearly cubic i.e. the effect of $\delta'$ is very small. 
    This justifies for \bairo\ the so-called T-P approximation \cite{Sugano} or $j_\textrm{eff}$-picture, commonly used for Ir-based oxides \cite{Kim1,Martins_2017}. 
    In the T-P approximation, initially introduced for cubic crystal fields when $\lambda \ll \Delta$, the $t_{2g}$ and $e_g$ blocks are treated separately and the SOC is applied to the $t_{2g}$ manifold only, giving rise to $j^{\text{cub}}_{\mathrm{eff}}$ eigenstates by diagonalizing $H_{t_{2g}}^{\text{loc}}$.
    Given the $d^5$ configuration in \bairo , this scheme thereby motivates on a model level the use of a three-band description. 

    Within the T-P approximation, the consequences of the \textit{tetragonal} crystal field on the eigenspectrum are also shown in  \autoref{fig:LevelDiag}. 
    The main difference with respect to the cubic case is that the \JEFF\ states are no longer degenerate leading to three doubly degenerate $j_{\mathrm{eff}}$ states.
    In addition, the presence of finite $\delta$ implies a certain degree of deviation in the eigenstates with respect to the ones relative to a cubic field.
     For a general tetragonal field, we can express the $j_{\mathrm{eff}}$ states as: 
    \begin{align*}
        \left| \frac{1}{2}, \pm\frac{1}{2} \right\rangle &= \frac{\sin\theta}{\sqrt{2}}\left( \left| d_{yz}\mp\right\rangle \pm i \left\vert d_{xz}\mp\right\rangle\right) \mp \cos\theta \left\vert xy\pm\right\rangle\\ 
        \left| \frac{3}{2}, \pm\frac{1}{2} \right\rangle &= \frac{\cos\theta}{\sqrt{2}}\left( \pm\left| d_{yz}\mp\right\rangle + i \left\vert d_{xz}\mp\right\rangle\right) + \sin\theta \left\vert xy\pm\right\rangle\\ 
        \left| \frac{3}{2}, \frac{3}{2} \right\rangle &= \frac{1}{\sqrt{2}}\left( \mp \left \vert d_{yz}\pm\right\rangle - i\left\vert d_{xz}\pm\right\rangle\right),\numberthis
        \label{eq:jeffgeneral}
    \end{align*}
    where we still denote (up) down spins with ($+$) $-$ and the angle $\theta$ is determined from $\delta$ and $\lambda$ .
    In particular, the cubic $j^{\mathrm{cub}}_{\mathrm{eff}}$ states are obtained for $\cos\theta=1/\sqrt{3}$, giving equal contributions of all three $t_{2g}$ orbitals to the \jeff\ state. 
    Instead, in \bairo\ the finite value of $\delta$ leads to $\cos\theta \approx0.7/\sqrt{3}$. 
    
    The departure from the cubic value of
    $\cos\theta$ induces a prominent $d_{xy}$ character of the $ \left| 3/2, \pm 1/2 \right\rangle$ state as well as the uneven contribution of the $t_{2g}$ orbitals to $ \left| 1/2,\pm 1/2 \right\rangle$.    
    \begin{figure}[tb!]
    \centering
    \includegraphics[width=.95\linewidth]{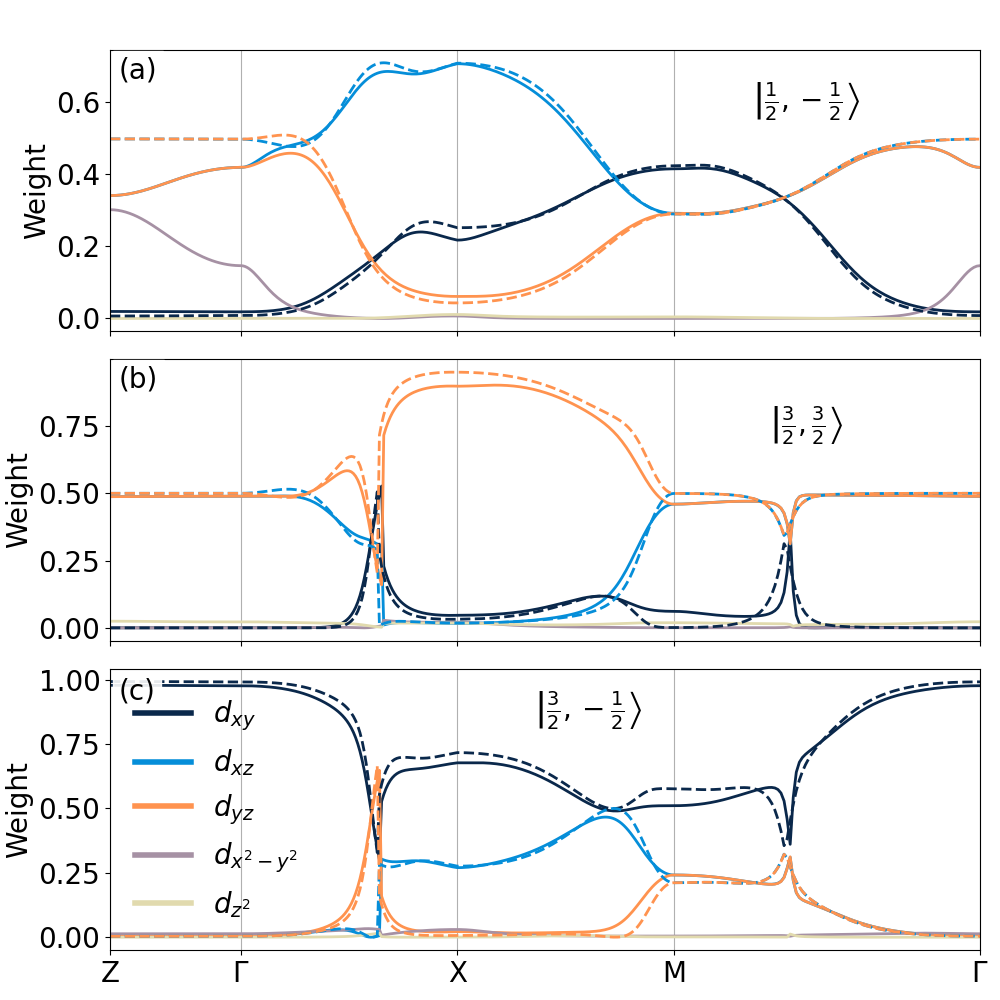}
    \caption{
    $\mathbf{k}$-resolved Wannier decomposition of the $j_{\mathrm{eff}}$ states.
    For the five-band model (full lines) and the three-band model (dashed lines), the $j_{\mathrm{eff}}$ states are decomposed in terms of the Wannier orbitals of the Ir atoms. %
        }
    \label{fig:jeffCharacter}
    \end{figure}
   This can also be seen in \autoref{fig:jeffCharacter} (a-c), which displays the $ \mathbf{k}$-resolved composition of the $j_{\mathrm{eff}}$ states in terms of Wannier orbitals of the Ir atoms for the five-band  Ir-$5d$ model (solid lines) and the effective three-band model (dashed lines). 
    The two models nearly perfectly agree within the $xy$ plane.   %
    Along $\Gamma$-X-M ($\Gamma$-Y-M) the main contribution to \jeff\ is given by the $d_{xz}$ ($d_{yz}$) orbital.
    Due to their degeneracy, $d_{xz}$ and $d_{yz}$ equally contribute along the $\Gamma-M$ direction. 
    Except for the vicinity of $M$, the weight of the $d_{xy}$ orbital is generally smaller than the one carried by the other two orbitals.
    Similarly, the $ \left| 3/2,\pm 3/2 \right\rangle$ band is dominated by $d_{xz}$ and $d_{yz}$ contributions. 
    This non-trivial composition of the pseudospin manifold has already been been observed in iridates  and similar materials \cite{Arpes_Sr2IrO4,Martins_SrIrO4_vs_SrRhO4} and is related to the crystal-field anisotropy. 
   We finally note that the $e_g$ states carry very small weight on the $j_{\mathrm{eff}}$ pseudospin states, except for the non-negligible contribution along Z-$\Gamma$.
    There, we observe an increasing contribution of the $d_{x^2-y^2}$ orbital towards $Z$ whereas the one of $d_{xz}$/$d_{yz}$ diminishes. 
    Within the three-band model, the absence of hybridization with the $e_g$ states leads to a constant contribution of $d_{xz}$/$d_{yz}$ along the $Z-\Gamma$ path.\\

    We finally turn to the local Coulomb interactions evaluated using cRPA, whose values are reported in \autoref{tab:Coulomb_Ba2IrO4} and \autoref{tab:Coulomb5b}. Note that in contrast to a recent study on Ca$_5$Ir$_3$O$_{12}$ \cite{Charlebois21}, we here calculate the cRPA interactions for the Wannier functions without SOC.
       \begin{table}[tb]
        \begin{minipage}{0.45\linewidth}
        \centering
        \begin{tabular}{c|ccc}
            \hline
            \hline
             U (eV) & $d_{xy}$ & $d_{yz}$ & $d_{xz}$ \\
             \hline
             $d_{xy}$ & 2.49 & 1.81 & 1.81 \\
             $d_{yz}$ & 1.81 & 2.39 & 1.86 \\
             $d_{xz}$ & 1.81 & 1.86 & 2.39 \\
            \hline
            \hline
        \end{tabular}
        \end{minipage}
        \begin{minipage}{0.45\linewidth}
        \centering
        \begin{tabular}{c|ccc}
            \hline
            \hline
             J (eV) & $d_{xy}$ & $d_{yz}$ & $d_{xz}$ \\
             \hline
             $d_{xy}$ & 0.00 & 0.22 & 0.22 \\
             $d_{yz}$ & 0.22 & 0.00 & 0.22 \\
             $d_{z}$ & 0.22 & 0.22 & 0.00 \\
            \hline
            \hline
        \end{tabular}
        \end{minipage}
        \caption{Coulomb and exchange parameters obtained for the three-band model of \bairo.}
        \label{tab:Coulomb_Ba2IrO4}
    \end{table}

    \begin{table}[tb]
        \centering
        \begin{tabular}{c|ccccc}
        \hline
        \hline
        U (eV) & $d_{xy}$ & $d_{yz}$ & $d_{xz}$ & $d_{x^2-y^2}$ & $d_{z^2}$ \\
        \hline
        $d_{xy}$ & 2.41 & 1.73 & 1.73 & 2.16 & 1.73 \\
        $d_{yz}$ & 1.73 & 2.31 & 1.78 & 1.83 & 2.01 \\
        $d_{xz}$ & 1.73 & 1.78 & 2.31 & 1.83 & 2.01 \\ 
        $d_{x^2-y^2}$ & 2.16 & 1.83 & 1.83 & 2.81 & 1.86 \\
        $d_{z^2}$ & 1.73 & 2.01 & 2.01 & 1.86 & 2.64 \\
        \hline
        \hline
        \end{tabular}

        \begin{tabular}{c|ccccc}
        \hline
        \hline
        J (eV) & $d_{xy}$ & $d_{yz}$ & $d_{xz}$ & $d_{x^2-y^2}$ & $d_{z^2}$ \\
        \hline
        $d_{xy}$ & 0.00 & 0.22 & 0.22 & 0.22 & 0.27 \\
        $d_{yz}$ &  0.22 & 0.00 & 0.22 & 0.26 & 0.21\\
        $d_{xz}$ &  0.22 & 0.22 & 0.00 & 0.26 & 0.21 \\ 
        $d_{x^2-y^2}$ &  0.22 & 0.26 & 0.26 & 0.00 & 0.36\\
        $d_{z^2}$ & 0.27 & 0.21 & 0.21 & 0.36 & 0.00 \\
        \hline
        \hline
        \end{tabular}
        \caption{Coulomb and exchange parameters obtained for the five-band model of \bairo}
        \label{tab:Coulomb5b}
    \end{table}
    
    Comparing the Coulomb parameters obtained within the five-band model and the three-band model, we observe that the $t_{2g}$ subspace of the five-band model has $\sim0.08$ eV smaller values. 
    The Hund's coupling $J$ within the $t_{2g}$ manifold is identical in the two models.
    On the other hand, the $t_{2g}-e_g$ part of the Coulomb matrix can be substantial.
    In particular the $d_{xy}-d_{x^2-y^2}$ and the  $d_{xz/yz}-d_{z^2}$ interactions are greater than $2$ eV. 
    We note that the Coulomb matrices reflect that different Wannier functions are used in the two models: 
    If the Wannier functions were the same, the larger screening via the $e_g$ bands would lead to much smaller values for the interactions within the three-band model, which is not the case. 
    
    We finally stress that in both cases the \textit{ab initio} calculated Coulomb tensor deviates from a spherical Hubbard-Kanamori representation \cite{Sugano}, especially due to the anisotropy of the density-density terms, which must be seen as another footprint of the non-trivial crystal field for this compound.
    
\section{Five-band vs. three-band model: A DMFT-based comparison}
\label{section:validation}

In this section, we present results of paramagnetic DMFT calculations for the five-band model of Ba$_2$IrO$_4$ and compare them to results obtained for the three-band model at the inverse temperature of $\beta=80$ eV$^{-1}$ ($T=145$ K). 
For the calculations, we use the continuous-time quantum Monte Carlo solver (CT-QMC) in the hybridization expansion matrix formulation (CT-HYB) \cite{TRIQS2015,TRIQSCTHYB2016,TRIQSDFTTOOLS2016} in order to keep all the off diagonal terms in the Hamiltonian resulting from inter-site mixing.
Other computational details of the calculations can be found in Appendix~\ref{appendix:Computational_details}.

    \begin{figure*}
        \centering
        \includegraphics[width=0.95\linewidth]{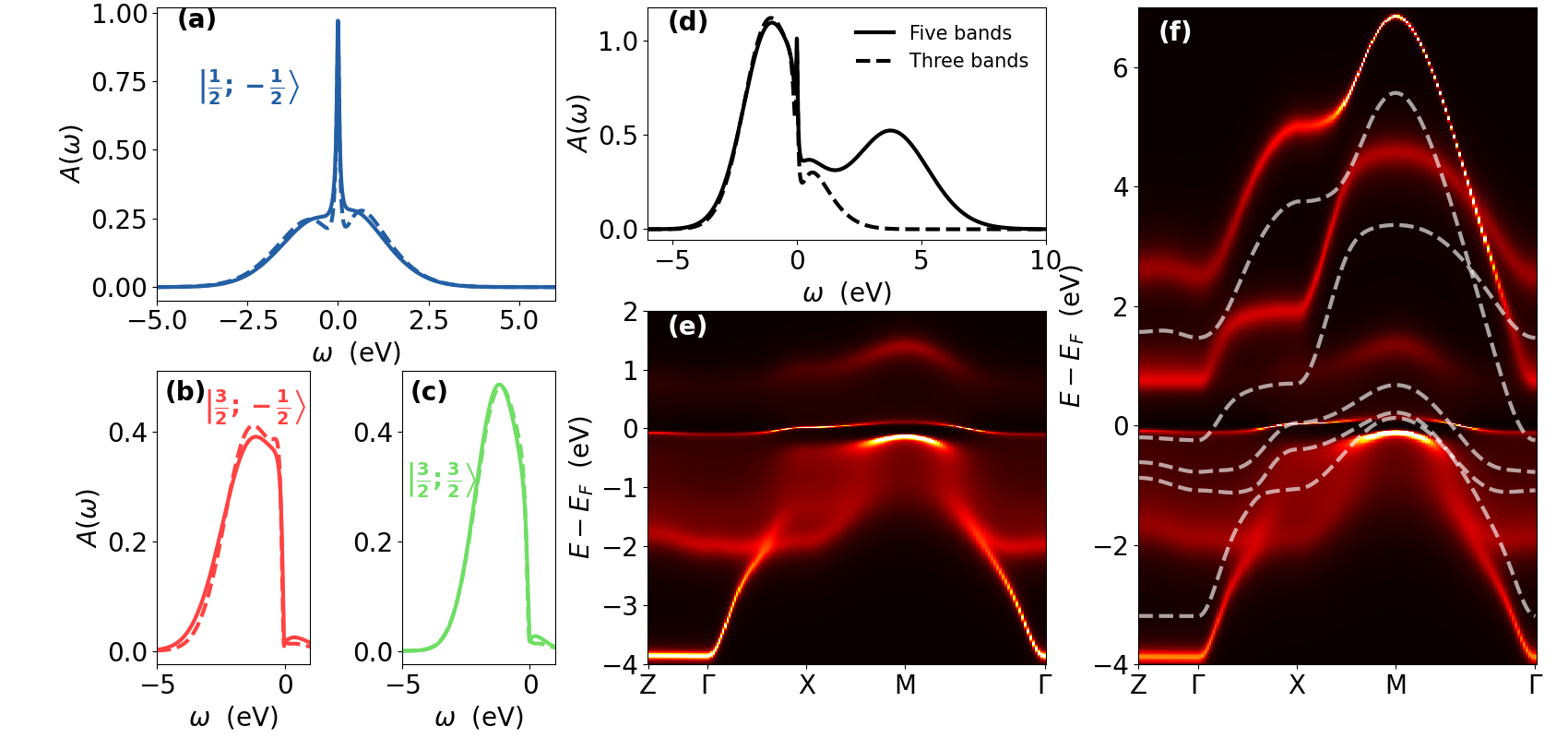}
        \caption{
        Comparison of the results of the three- and five-band DMFT calculations. 
        Contributions of the $\left| \frac{1}{2}; - \frac{1}{2} \right\rangle$ (a), $\left| \frac{3}{2}; - \frac{1}{2} \right\rangle$ (b) and $\left| \frac{1}{2}; -\frac{3}{2} \right\rangle$ (c) states to the total density of states $A(\omega)$, which is shown in (d), for the five-band (solid line) and three-band (dashed line) models. 
        The \textbf{k}-resolved spectral function $A(\mathbf{k},\omega)$ for the three-band (e) and five-band (f) models is plotted along the high-symmetry path $(0,0,\frac{\pi}{c})-(0,0,0)-(\frac{\pi}{a},0,0)-(\frac{\pi}{a},\frac{\pi}{a},0)-(0,0,0)$.
        }
        \label{fig:5_vs_3_akw}
    \end{figure*}
    
\begin{figure}
        \centering
        \includegraphics[width=0.9\linewidth]{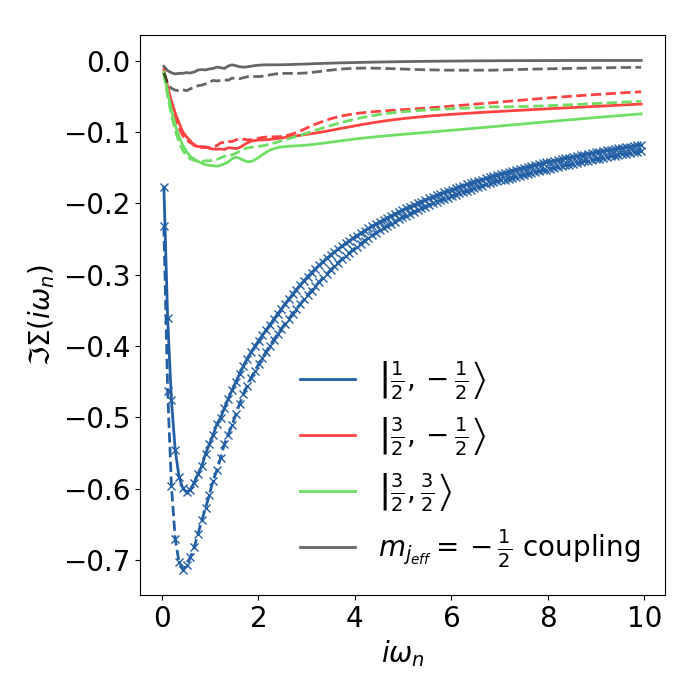}
        \caption{Imaginary part of the self-energy in Matsubara space. 
        Comparison of the $j_{\mathrm{eff}}$ components of the five-band (solid lines) and three-band (dashed lines) models shows that the three-band model is closer to the phase transition.
        } 
        \label{fig:self_energy}
    \end{figure}

    \autoref{fig:5_vs_3_akw} (a)-(d) compares the density of states of the two models, expressed in terms of the $j_{\textrm{eff}}$ components,
    which turn out to be in very good agreement in the electronic part of the spectrum.
    The three peak structure of the \jeff\ band is recovered in both models with the three-band model \jeff\ spectral function being closer to the Mott transition: 
    lower and upper Hubbard band are better defined and the quasiparticle peak is narrower than for the five-band model. 
    This is corroborated by the self energy shown in \autoref{fig:self_energy}.
    The imaginary part of the \jeff\ self-energy shows already a divergent behavior for $i\omega_n\rightarrow 0$ and only has an upturn for the lowest six Matsubara frequencies $\omega_n$. 
    The quasiparticle mass of the band, ${m^*}/{m}=Z^{-1}$, is larger for the three-band model than for the five-band model and the corresponding quasiparticle residue $Z=\left[1-\partial_\omega\Re \Sigma(\omega+i0^+)\vert_{\omega=0}\right]^{-1}\approx\left[ 1-\Im\Sigma(i\omega_0)/\omega_0\right]^{-1}$ is $Z=0.14$ ($Z=0.18$) respectively.

    The filled \JEFF\ bands also exhibit a remarkable agreement below the Fermi level, see \autoref{fig:5_vs_3_akw} (b) and (c). 
    Above the Fermi level, the \JEFF\ states of the five-band model exhibit some weight due to the hybridization with the $e_g$ states, which is absent in the three-band calculation. 
    As for the \jeff\ band, we observe a stronger renormalization due to the self-energy, see \autoref{fig:self_energy}, where the bandwidth renormalization is 0.69 (0.71) for the $m_{j_{\mathrm{eff}}} = \frac{3}{2}$ component and 0.72 (0.76) for the $m_{j_{\mathrm{eff}}} = - \frac{1}{2}$ component of the three-band (five-band) model.
    The \textbf{k}-resolved spectral functions plotted in \autoref{fig:5_vs_3_akw} (d) and (e) show remarkably good agreement. 
    All features in the $j_{\mathrm{eff}}$ manifold are comparable, due to very similar $\Sigma(i\omega_n)$ and nearly identical TB models in the two cases. 
    In particular, we observe that the \JEFF\ states are completely filled, consistently with panels (b,c) and a strongly renormalized flat $j_{\mathrm{eff}}=1/2$ band appears at the Fermi level. 
    Compared to the initial DFT simulations, the $\tilde{e}_g$ bands are moved up above the Fermi level and do not cross it anymore.  
    This amounts nearly exclusively to a Hartree shift, additional effects of correlations and spin-orbit coupling on these bands are minimal, as for instance shown by the very small broadening of the bands. %
    This result, although expected, has to our best knowledge not been achieved before and is in agreement with available experiments~\cite{Moser, RIXS-Ba2IrO4}.

    \begin{table}[ht]
        \centering
        \begin{tabular}{|l|c|c||l|c|c|}
            \hline
            $j_{\mathrm{eff}}$ & 3 band & 5 band & $d$ & 3 band & 5 band \\
            \hline
            \hline
            $ \left| 3/2,1/2  \right\rangle$ & 0.987 & 0.985 & $\left| d_{xz} \right\rangle$ & 0.814 & 0.803\\
            $ \left| 3/2,3/2 \right\rangle$ & 0.990 & 0.989 & $\left| d_{yz} \right\rangle$ & 0.814 & 0.803\\
            $ \left| 1/2, 1/2 \right\rangle$ & 0.519 & 0.517 & $\left| d_{xy} \right\rangle$ & 0.868 & 0.856 \\
            $ \left| \tilde{d}_{x^2-y^2}  \right\rangle$& / & 0.003& $\left| d_{x^2-y^2} \right\rangle$ & / & 0.020\\
            $ \left| \tilde{d}_{z^2}  \right\rangle$& / & 0.006& $\left| d_{z^2} \right\rangle$ & / & 0.017 \\
            \hline
        \end{tabular}
        \caption{
        Band fillings within DMFT.
        For both models, the fillings are reported with respect to the $j_{\mathrm{eff}}$ basis (left) and the orbital basis (right) respectively.
        The calculations used cRPA values for the Coulomb tensor and $\beta = 80$ eV$^{-1}$. }

        \label{tab:filcrpa80}
    \end{table}
    To further quantify the agreement of the two models, we report in \autoref{tab:filcrpa80} the band fillings obtained from DMFT.
    The fillings of the $j_{\mathrm{eff}}$ bands are nearly the same for both models. 
    The two $j_{\mathrm{eff}}=3/2$ bands are slightly less filled in presence of $\tilde{e}_g$ orbitals. 
    In the $d$ basis, this translates to a slightly less filled $d_{xy}$ orbital which transfers some of its electrons to the $e_g$ orbitals. 
    This charge transfer is made possible due to the small coupling between the $t_{2g}$ and $e_g$ subspaces when adding SOC. 
    We note, however, that these fillings disagree with a previous RIXS and XAS study \cite{RIXS-Ba2IrO4} where the authors measured a larger hole filling in the $d_{xy}$ than in the $d_{xz}, d_{yz}$ states.
    Our results are though in qualitative agreement with previous theoretical simulations \cite{katukuri2012, katukuri2015}.
   To summarize, both models agree to a large extent in the ${j}_{\mathrm{eff}}$ sector.
    Given the non-negligible interactions in the $t_{2g}-e_g$ sector of the Coulomb matrix, this is a non trivial result.
    The rationale behind it is that the $e_g$ bands are essentially empty within our DMFT simulations, thus they do not have a notable impact on the band structure of the $j_{\mathrm{eff}}$ manifold.
    Our comparison thereby validates the applicability of the three-band model for the low-energy description of Ba$_2$IrO$_4$. 
    From a computational point of view this is an important result since the three-band model is at least one order of magnitude faster to solve numerically with DMFT.
    We will focus on it for the rest of the paper.\\

    The metallic solutions using the \textit{ab initio} computed parameters are in contrast to experiment and might at first glance seem surprising.
    However, it indicates primarily an underestimation of the Coulomb tensor.
    This is a well-known problem within cRPA, which tends to over-screen the Coulomb interaction~\cite{cRPA2}. 
    A partial solution to this problem was found by treating the dynamic part of $U(\omega)$, accessible by cRPA, whose static limit gives the cRPA Coulomb tensor. 
    It has been shown that the static effective model obtained by taking into account the frequency dependent $U(\omega)$ via a Lang-Firsov transformation is more correlated than the one including only the static limit of $U$\cite{Casula2012_PRB}. 
    The additional correlation comes from an effective bandwidth reduction, provided by the $Z_{LF}$ renormalization factor and estimated by the Lang-Firsov approximation\cite{Casula2012_PRL}. 
    We computed the full frequency dependence of the monopole term in cRPA for the 3-band model, and we found that the resulting bandwidth reduction factor is $Z_{LF}\approx0.84$. 
    This leads to an absolute increase in $U_{\mathrm{cRPA}}$ by a $\Delta U \approx 0.4 - 0.5$ eV.
    As we show in the following, such an increase already leads to an insulating solution within DMFT.\\

    In the next Section, we are going to study the effect of progressively increasing the interaction strength above the cRPA values on the resulting phase diagram, and we will evaluate it as a function of $\Delta U$, $\lambda$, $J$ and $T$ to study whether the effective model can be further reduced to a single-band \jeff\ model.

\section{Spin-orbit Coupling, Hund's exchange and the Mott transition}
\label{section:spin_orbit}

In this Section, we study the \textit{ab initio} three-band model as a function of interaction strength $U$, spin-orbit coupling $\lambda$, and Hund's exchange $J$ within DMFT. 
Computational details are given in Appendix~\ref{appendix:Computational_details}.
We will discuss the full phase diagram in order to carve out the nature of the Mott transition in Ba$_2$IrO$_4$.

\begin{figure}[tb!]
\centering
\includegraphics[width=0.95\linewidth]{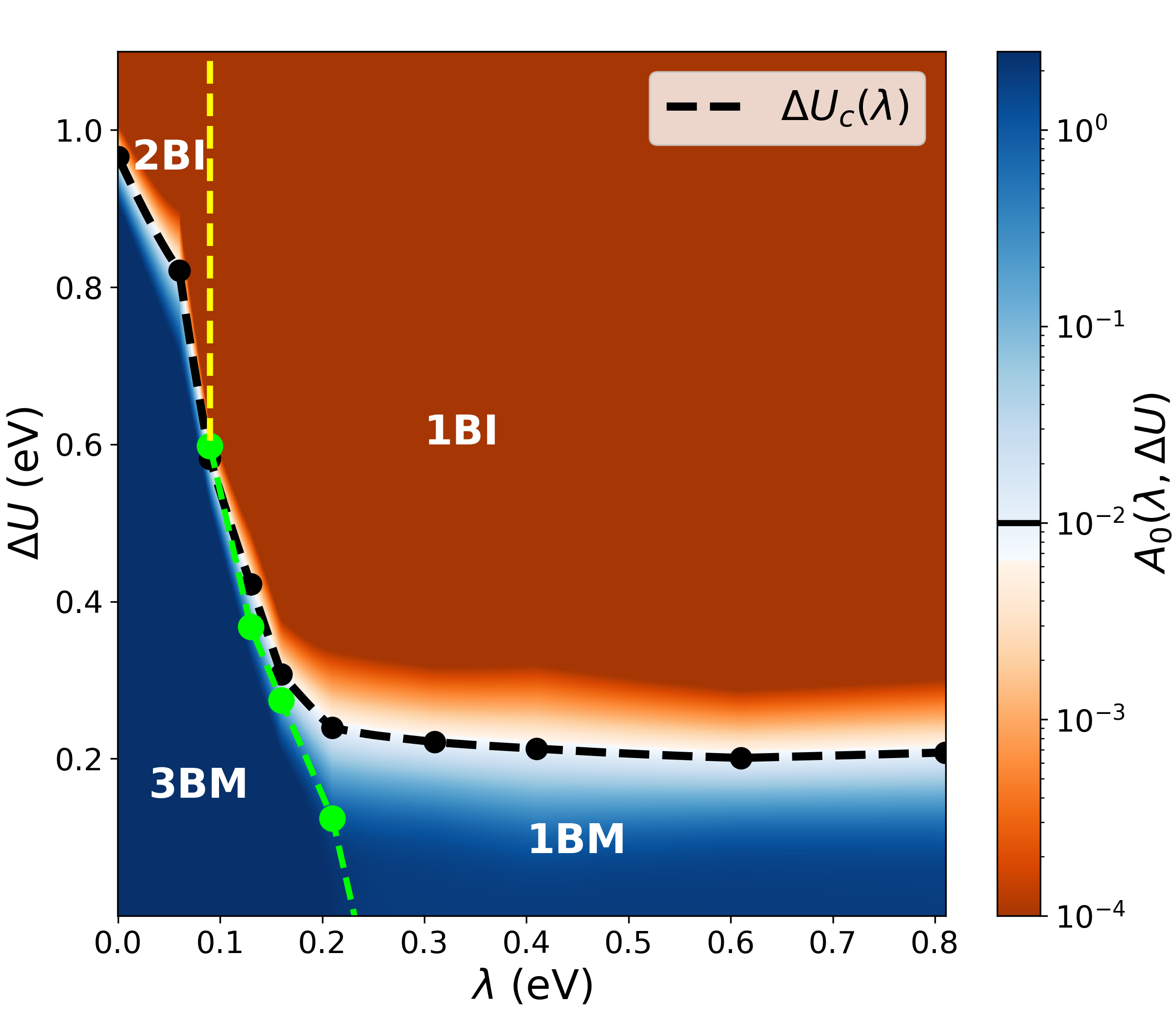}
\caption{ Metal-insulator transition in the three-band model of Ba$_2$IrO$_4$. 
The $U$ vs. $\lambda$ phase diagram at $T=193.5$ K with a Hund's coupling $J=0.22$~eV is defined by the density of states at the Fermi level,  $A_0(\Delta U, \lambda)$. %
1-band (3-band) metal solutions are denoted by 1BM (3BM) and separated along the transition line $\Delta U_c(\lambda)$ from the insulating solutions.
We denote insulators with one (two) of the bands split in lower and upper Hubbard band by 1BI (2BI).
We use a threshold value of $A_0\equiv10^{-2}\ \text{eV}^{-1}$ to discriminate metallic from insulating solutions.
The dashed green line indicates a Lifshitz transition in the metallic phase. 
}
\label{fig:SOC_diagrams}
\end{figure}

In order to calculate the phase diagram, we first increase the density-density part of the Coulomb tensor by an isotropic amount $\Delta U$ at a fixed temperature of $T=193.5$ K ($\beta=60$ eV$^{-1}$). 
In terms of the traditional representation through Slater integrals~\cite{slater}, this corresponds to increasing the monopole term $F_0$, leaving the higher-order terms unchanged. 
In absolute values, the critical transition line that marks the metal-insulator transition (MIT) is expressed as $U_c(\lambda)= U_{\mathrm{cRPA}}+\Delta U_c(\lambda)$.  
To distinguish the metallic phase from the insulating one, we analyze 
the density of states (DOS) at the Fermi level and its evolution within the parameter space.
The DOS value $A_0(\Delta U, \lambda)$ as a function of $\Delta U$ and $\lambda$, by keeping $J$ equal to its cRPA value (Tab.~\ref{tab:Coulomb_Ba2IrO4}), determines the phase diagram reported in  \autoref{fig:SOC_diagrams}.
We should note that it is calculated for a temperature above the second-order end point of the Mott transition.
Therefore, for fixed $\lambda$ a crossover from a bad metal to a bad insulator is realized as a function of $U$. 
The dashed line $\Delta U_c(\lambda)$ serves as a guide to the eye to identify a good insulator.
In particular, we chose a threshold value for $A_0$, such that $\Delta U_c(\lambda)$ is defined by $A_0(\Delta U_c,\lambda)\equiv 10^{-2}$ eV$^{-1}$.
Further details on the interpolation scheme used for plotting $A_0$ are reported in Appendix~\ref{appendix:interpolation_phase_diagrams}.

By inspection of the orbital-resolved spectral functions, four distinct regions can be defined: 
a three-band metal (3BM), a single-band metal (1BM), a two-band insulator (2BI), and a single-band insulator (1BI). 
For $\lambda\neq0$ this characterization is done in the ${j}_{\mathrm{eff}}$ basis.
We furthermore indicate in green a SOC-induced Lifshitz transition \cite{Lifshitz}, which transforms the system from 3BM to 1BM and vice versa, i.e. between metals with a three-sheet and a one-sheet Fermi surface, respectively. 
Most interestingly, the overall phase diagram is qualitatively similar to the one of a spin-orbit coupled, quarter-filled two-band Hubbard model reported in Ref.~\onlinecite{Enhanced_SOC_correl}.\\

The remainder of this section is divided in four Subsections.
First, we discuss the MIT line and the topological Lifshitz transition (Secs.~\ref{subsection:SOC_and_MIT} and \ref{subsection:SOC_and_Lishfits}). 
Afterwards, Sec.~\ref{subsection:Hund} is dedicated to the impact of Hund's coupling on the MIT.
Finally, fixing the SOC constant to its \textit{ab initio} value of $\lambda=0.31$ eV, we will derive a $U-T$ phase diagram for \bairo\ (Sec.~\ref{subsection:ut_phase_diagram}).
      \subsection{The metal insulator transition}
      \label{subsection:SOC_and_MIT}
      Analyzing the behavior of $\Delta U_c(\lambda)$ with respect to the spin-orbit coupling strength, we can distinguish three regions: weak, intermediate and strong SOC. \\

\begin{figure}[bt!]
\centering
\includegraphics[width=.95\linewidth]{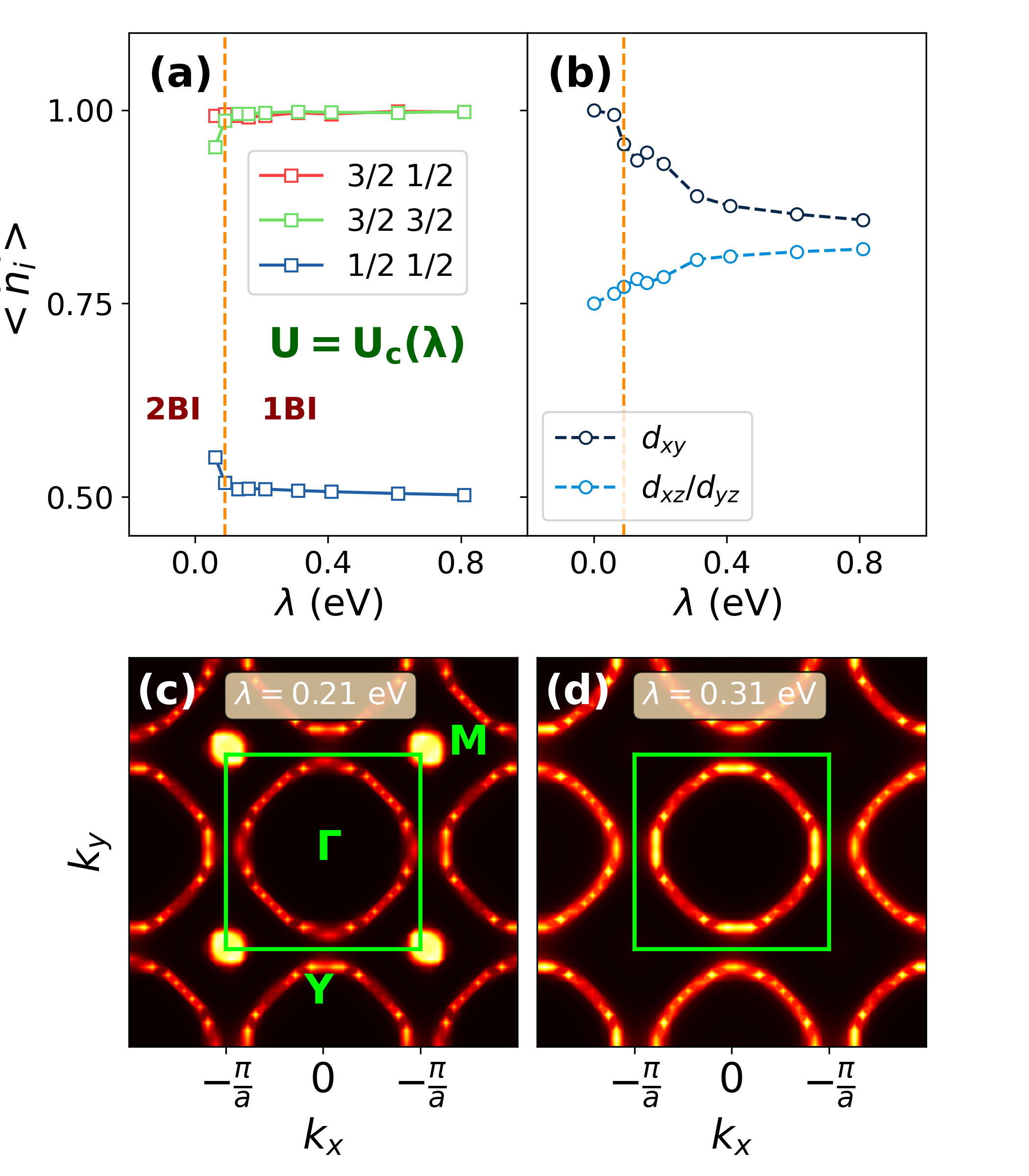}
\caption{ Characteristics of the metallic and insulating phases.
The diagonal density-matrix elements (filling) of the insulating state close to the transition line, i.e. along $\Delta U_c (\lambda)$, are expressed in the $j_\mathrm{eff}$ (a) and in the $t_{2g}$ (b) basis. 
The Fermi surface of the model system at $\Delta U=0.0$ eV (i.e. $U=U_{\text{cRPA}}$) is shown for $\lambda=0.21$ eV (c) and for $\lambda=0.31$ eV (d). 
}
\label{fig:SOC_diagrams_details}
\end{figure}

      The weak SOC region is defined for $\lambda \lesssim 0.1$ eV. 
      There, small variations of the coupling constant massively decrease the value of $\Delta U_c$. 
      Moreover, the MIT is very sharp.
      We gain further insights from looking at the diagonal elements of the density matrix of the insulating state for $U\sim U_c$, see \autoref{fig:SOC_diagrams_details} (a) and (b).
      Excluding the completely filled bands, we obtain either one or two partially filled bands that are split by the electron-electron interactions into upper and lower Hubbard bands (UHB/LHB).
      In other words, the 2BI phase is realized when we only have one band completely filled, while the 1BI is found when two bands are filled.
      Consistently with the indicated dashed line in the phase diagram, we find a 2BI for $\lambda \lesssim 0.1$ eV and strong interactions, $U>U_c(\lambda)$.

      The sharpness of the metal-insulator transition can be explained as follows. 
      Since the SOC is small, a representation in terms of $t_{2g}$ orbitals should still be a good description.
      The phase transition occurs when the interaction strength is large enough to move one band below the Fermi level via the relative Hartree shift.
      Being the one with lower local energy, this turns out to be the band of dominant $d_{xy}$ character. 
      At the non-interacting  Kohn-Sham level, the difference in local energies $\delta$ is compensated by larger inter-site hoppings in the $xy$ plane that yield nearly isotropic filling.
      In DMFT, the band anisotropy determines different Hartree shifts $\Delta_{xy}<\Delta_{xz}=\Delta_{yz}$, thus effectively enhancing the difference in the local energies, until the band $d_{xy}$ becomes completely filled.
      Once this occurs, the system can be effectively described by a two-band model with an average filling of $\langle n\rangle \approx 3/4$ in the $t_{2g}$ representation. 
      In that frame, the amount of correlation is already sufficient to realize a Mott insulating state for the $d_{xz}$ and the $d_{yz}$ bands.\\
      In the opposite limit of large SOC, $\lambda\gtrsim0.6$ eV, we observe a different situation. 
      There, the critical line $U_c(\lambda)$ is flat,  nearly independent of $\lambda$. 
      We can reveal the reason for this behavior by looking at the model in the non-interacting limit. 
      For $\lambda\gtrsim0.6$ eV the ${j}_{\mathrm{eff}}=1/2$ band is half-filled, thereby being the only one crossing the Fermi energy. 
      Due to strong SOC, the $j_{\mathrm{eff}}=3/2$ states are already completely filled by construction, even without interaction.
      Still, we will refer to this situation as an \textit{effective} single-band problem, since there is a non negligible mixing in the non-local part of the TB Hamiltonian between the $ \left| 3/2;1/2  \right\rangle$ and $ \left| 1/2;1/2  \right\rangle$ states.
      Thus, we are in an \textit{effective} single-band ${j}_{\mathrm{eff}}=1/2$ problem at half-filling, whose Mott transition is entirely driven by interactions.
      We finally note that the critical value of $\lambda_c$ to realize this effective single-band problem is in quantitative agreement with previous studies for similar model systems \cite{Bethe_SOC_all_filled,Shinauka_multipole}.
      Any further increase in $\lambda>\lambda_c$ does not change this situation qualitatively, which is why the critical interaction strength $U_c$ is only mildly affected. 
      Also this result turns out to be qualitatively in line with previous studies on the Bethe lattice in the strong SOC regime \cite{Shinauka_multipole}.  
      Since the  ${j}_{\mathrm{eff}}=3/2$ bands are already below the Fermi level due to strong SOC, the critical interaction strengths are lower than in the case of weak SOC.\\ 

      In the intermediate range, for $0.1 \lesssim \lambda \lesssim 0.6$ eV, the insulating state still involves a single half-filled band, but a finite interaction strength is needed to push the ${j}_{\mathrm{eff}}=3/2$ band below the Fermi level via a Hartree shift. 
      The consequences of this are particularly evident in the range  $0.1 \lesssim \lambda \lesssim 0.2$ where $\Delta U_c(\lambda)$ develops a slope similar to the one in the weak SOC regime and, on the metallic side, the transition goes directly from a 3BM to a 1BM by increasing $\Delta U$.\\

      Given this analysis we can draw some partial conclusions. 
      (i) As a consequence of the orbital polarization induced by the ligand field $\delta$, the transition is always orbital selective. 
      (ii) Due to the orbital polarization, the transition occurs over the entire phase diagram for values of $U$ which are moderate, considering the electronic filling and the number of orbitals involved. 
      (iii) In the strong SOC regime, we have an effective single-band problem, which justifies the construction of $j_{\mathrm{eff}}=1/2$ models. 
      For intermediate SOC, however, such an effective single-band model is only valid close to the metal-insulator transition.\\

     Finally, \autoref{fig:SOC_diagrams_details} (b) also provides some insights with respect to the natural basis to formulate the problem. 
     Most importantly, a single-orbital, selective transition in the ${j_{\mathrm{eff}}}$ basis is a pure Mott transition involving all states from the $t_{2g}$ viewpoint. 
     Moreover, upon increasing $\lambda$ the entanglement of the $t_{2g}$ orbitals increases and the electronic distribution is reshaped within the orbital manifold, potentially leading to a perfectly isotropic $j_{\mathrm{eff}}=1/2$ state in terms of the $t_{2g}$ orbitals in the limit $\lambda \rightarrow  \infty$.

\subsection{Spin-orbit induced Lifshitz transition}
      \label{subsection:SOC_and_Lishfits}

     Besides the MIT, we investigate now the nature of the metallic phase, distinguishing the two topologically different regions 3BM and 1BM. 
     The dotted green line in \autoref{fig:SOC_diagrams} marks this SOC-driven Lifshitz transition \cite{Lifshitz}. 
     At $\Delta  U=0$ eV the transition occurs at $0.2 < \lambda < 0.3$ eV as shown in \autoref{fig:SOC_diagrams_details} (c) and (d). 
     For $\lambda=0.21$ eV, we have a clear ${j}_{\mathrm{eff}}=1/2$ sheet dispersing around the $\Gamma$ point, but we can recognize as well poles coming from the two ${j}_{\mathrm{eff}}=3/2$ bands at $M=(\pi/a,\pi/a,0)$. 
     Those poles disappear for $\lambda=0.31$ eV, consistently with the electronic structure shown in \autoref{fig:5_vs_3_akw} (e). 
     Interestingly, for $0.09 \leq \lambda \leq 0.16$ eV, the Lifshitz transition line merges with the metal-insulator transition line. 
     In that region, the transition is the narrowest of the entire phase diagram and we pass from a 3BM to a 1BI. 
     By slightly tuning $\lambda$ and $\Delta U $ in this section of the phase diagram, all phases of the system can be reached. 
     Despite the value of $\lambda$ being in the typical order of magnitude observed for heavy $4d$ compounds \cite{Martins_SrIrO4_vs_SrRhO4,Pavarini_Sr2RuO4_Anysotropy}, a real material with these properties does not exist so far.
     Nevertheless, it could show an intriguing and exotic variety of competing phenomena as a function of $\lambda$, $U$ and $\delta$. 
     However, decreasing the interactions will increase the value of $\lambda$ necessary to induce the change of topology. 
     Finally, at $\lambda=0.6$ eV the Lifshitz transition happens already in the non-interacting limit $U=0$ eV.\\ 

     In this perspective, the three regions identified by the dispersion of $\Delta U_c(\lambda)$ can be characterized in terms of a more general argument:
     The large SOC regime is where the single band problem is realized without any correlation by SOC. 
     The intermediate SOC region is where this mapping onto the single band problem in correspondence to the transition line occurs thanks to cooperation of spin-orbit coupling and interactions. 
     Ba$_2$IrO$_4$ and most of the iridium compounds of the Ruddlesden-Popper series are actually placed in this intermediate coupling region \cite{Moser,Electronic_structure_of_iridates}. 
     On the other hand, for the weak SOC regime, a single-band picture is not valid, not even close to $U_c$.
     Instead, within the intermediate and strong SOC region, a single-band paramagnetic Mott transition is found, induced by spin-orbit coupling.\\   

    \subsection{The Role of Hund's coupling}
      \label{subsection:Hund}

  Before investigating the impact of Hund's coupling  on the picture derived so far,
  we note that despite $J$ having been formally introduced as a tensor on the correlated manifold, the cRPA calculations yield the isotropic value of 0.22 eV on the full $t_{2g}$ subset. 
  Thereby, we can refer to it as a scalar with no ambiguity. \\

\begin{figure}
        \includegraphics[width=0.99\linewidth]{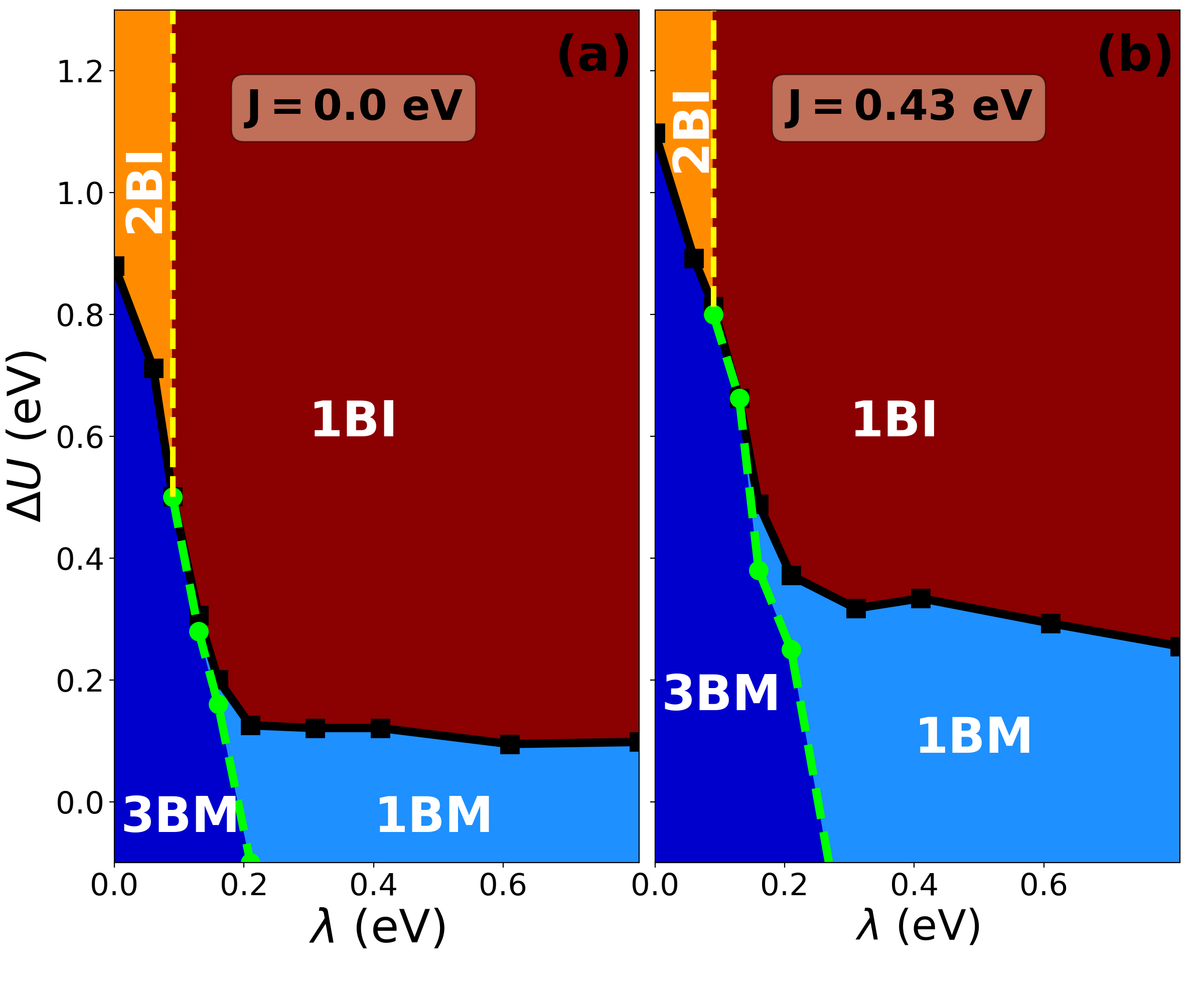}  
        \caption{Phase diagram for Hund's coupling $J=0.0$ eV in (a) and for $J=0.43$ eV in (b). 
        The phase diagrams are qualitatively similar to the one shown in \autoref{fig:SOC_diagrams} at  $J=0.22\ \text{eV}$.}
        \label{fig:SOC_vs_J}
    \end{figure}

  We  derived three $U-\lambda$ phase diagrams depending on the $J$ value, the first being the one with respect to the \textit{ab initio} value ($J=0.22$~eV) explained in detail in the previous sections.
  In addition, we considered a vanishing Hund's coupling and finally, we doubled the \textit{ab initio} value. 
  The resulting phase diagrams are shown in \autoref{fig:SOC_vs_J}. 
  In analogy to \autoref{fig:SOC_diagrams}, each of them is composed by the four regions 3BM, 1BM, 2BI, 1B1. 
  Based on $\lambda$, the three regimes identified for $J=0.22$~eV still hold and the dispersion of $\Delta U_c (\lambda)$ is qualitatively the same in all cases. 

     \begin{figure*}[t!]
    \begin{minipage}{\textwidth}
        \centering
        \includegraphics[width=0.99\linewidth]{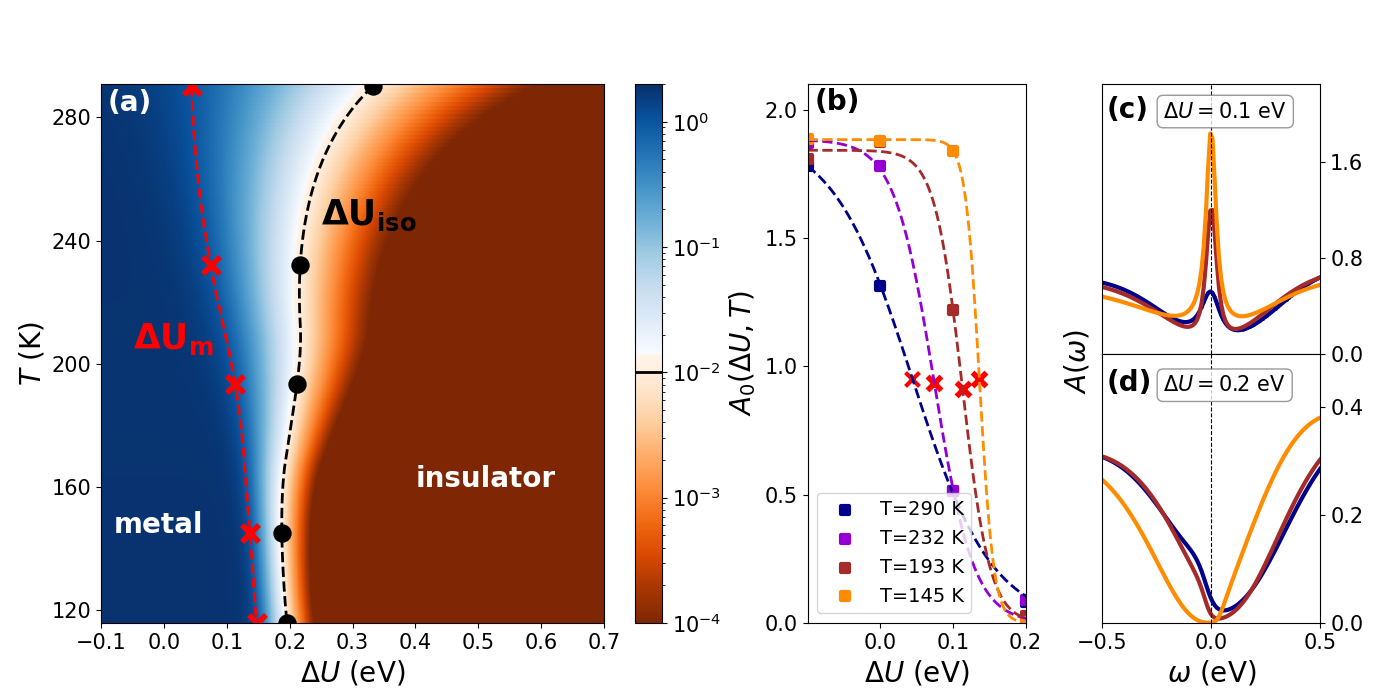}  
        \caption{Temperature-dependence of the metal-insulator transition.
        (a) U-T phase diagram for Ba$_2$IrO$_4$ with $\lambda=0.31$ eV and $J=0.22$ eV. 
        (b) Selected cuts at fixed temperature $T$ for $A_0(\Delta U,T)$ showing the metal-insulator transition (crossover) at low (high) temperature.
        Red crosses and black dots indicate the positions of the inflection point, $\Delta U_m$, and the iso-value line at $A_0\equiv10^{-2}\ \text{eV}$, $\Delta U_{\text{iso}}$, respectively.
        The evolution of the spectral function $A(\omega)$ projected on the \jeff\ state for different temperatures at $\Delta U=0.1$ eV (c) and $\Delta U =0.2$ eV (d). 
        The color schemes are the same as in (b). 
        }
   \label{fig:U-T-phase_diagram}
   \end{minipage}
   \end{figure*}
  
  Overall, increasing the Hund's exchange term delays both the MIT and the Lifshitz transition, and can be interpreted as a footprint of its competition with both $U$ and $\lambda$.
  Considering $J=0.43$ eV, our calculations reveal a larger intermediate 1BM metallic state between the 3BM and the 1BI for $\Delta U=0.3$ eV and $\lambda=0.16$ eV. 
  Thus, the transition line directly separating 3BM from 1BI shortens for larger $J$.
  For $\lambda=0.31$ eV, as in \bairo, the MIT occurs at $\Delta U=0.1$ eV ($0.3$ eV) for $J=0.0$ eV ($0.43$ eV) respectively. 
  The sensitivity of the phase diagrams with respect to $J$ shows that a multi-orbital treatment is required to describe the physics properly.
  In particular in the intermediate SOC regime, the coupling of different orbitals through the exchange term can be substantial. 
  This fact and the presence of non-negligible mixing terms between the different pseudospin states renders \bairo\ a multi-orbital compound in its essence.

\subsection{U-T phase diagram}    
\label{subsection:ut_phase_diagram}

    After having discussed the effects of spin-orbit coupling and exchange, we now focus on values relevant to \bairo, i.e. $\lambda=0.31$ eV and $J=0.22$ eV.
    \autoref{fig:U-T-phase_diagram}~(a) shows the corresponding phase diagram as a function of temperature and interaction strength $\Delta U$.
    We again quantify the metal-insulator transition by the value of the spectral function at the Fermi level, $A_0(\Delta U,T)$, now expressed as a function of correlation strength and temperature. 
    In analogy with \autoref{subsection:SOC_and_MIT}, the transition is generally broad, with a crossover region between the insulating and metallic phases. 
    We mark the middle of the crossover region, $U_m$, as well as the iso-value line of the spectral weight at the Fermi level, $\Delta U_{\text{iso}}=\Delta U_c(\lambda)$. 
    The latter can serve as a guide to identify the insulating phase and thereby an end of the crossover region, similar to $\Delta U_c$ in the previous section. 
    On the other hand, $\Delta U_m$ defines the inflection point of the spectral function with respect to $\Delta U$, $\partial^2_{\Delta U} A_0(\Delta U,T)|_{\Delta U_{m}}$=0. 
    Within the crossover region, the spectral weight at the Fermi level is suppressed with respect to the metallic phase, but it is still non-vanishing due to thermal excitations. 

    The size of the crossover region, and the reference values $\Delta U_m$, $\Delta U_{\text{iso}}$, depend on the temperature. 
    At high temperature, $290 \lesssim T \lesssim 190$ K, the smooth and continuous metal-insulator crossover extends over several hundreds of meV.  
    For lower temperatures, the crossover region gets narrower and turns into a sharp transition line at $T\lesssim150$ K.
    This is consistent with the picture of a second order end point of the first-order metal-insulator transition line known from the Mott transition \cite{Terletska2011,Vucicevic2013}.
    We report this temperature effect in more detail in \autoref{fig:U-T-phase_diagram}~(b), which displays the evolution of $A_0(\Delta U,T)$ at different temperatures. 
\begin{figure*}[t!]
 \begin{minipage}{\textwidth}
\centering
\includegraphics[width=0.99\linewidth]{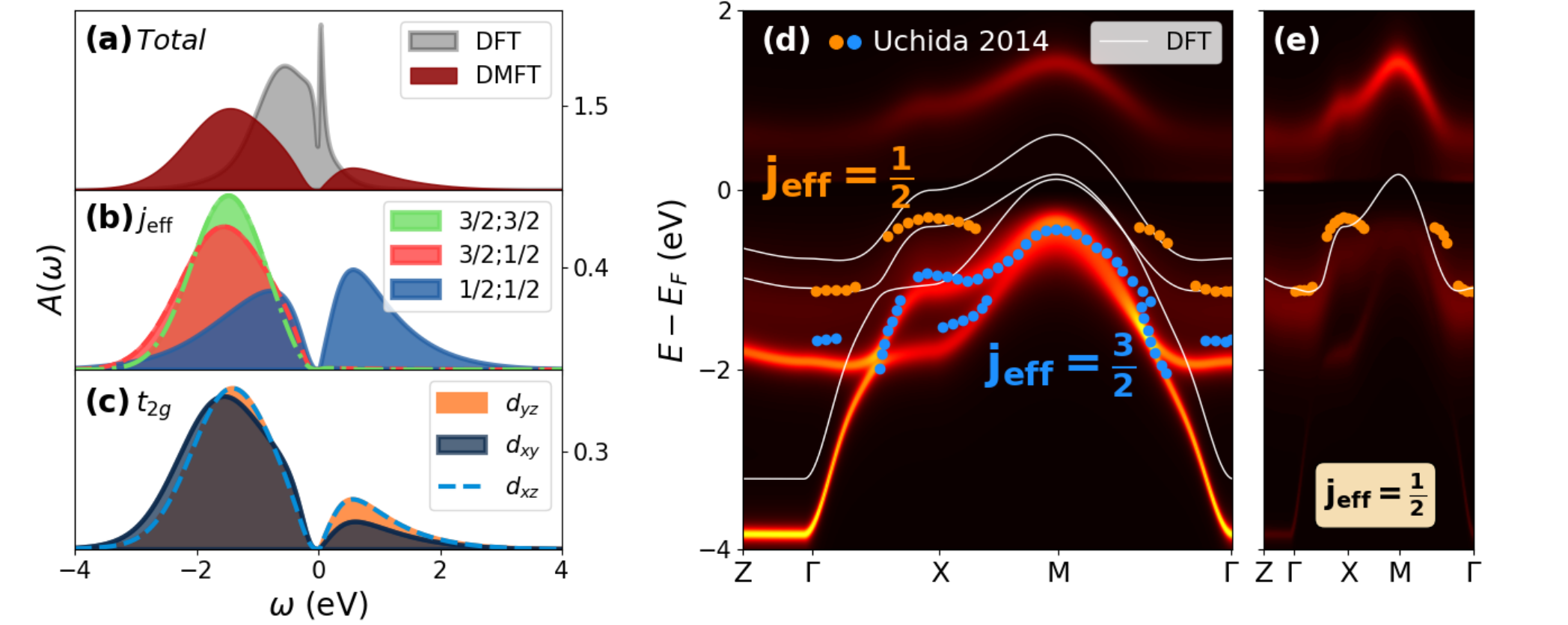}
\caption{ Spectral function of Ba$_2$IrO$_4$.
(a) Comparison between DFT and DMFT density of states as well as the orbital-resolved spectral function of the latter in the $j_{\mathrm{eff}}$ basis (b) and in the $t_{2g}$ basis (c).
The total \textbf{k}-resolved spectral function is shown in (d), its \jeff\ component in (e).
The DFT band structure of the $j_{\mathrm{eff}}$ bands is shown in white lines; orange and blue points correspond to intensity maxima of energy distribution curves extracted from the ARPES measurements of Ref.~\onlinecite{Uchida_ARPES}. 
All DMFT calculations were done for $\Delta U=0.2$ eV at $\beta=80\ \text{eV}^{-1}$.  
}
\label{fig:bands_and_cuts}
\end{minipage}
\end{figure*}
    Panels (c) and (d) show the evolution of the spectral function $A(\omega)$ for different temperatures, projected on the $j_{\mathrm{eff}}=1/2$ state for fixed $\Delta U$. 
    The quasi-particle peak for $\Delta U=0.1$ eV is more pronounced for lower temperature indicating the formation of a well-defined Drude peak in the metallic phase.
    Increasing the Coulomb interaction strength to $\Delta U=0.2$ eV, we observe the opposite behavior: 
    At $T=290$ K the system is a bad insulator, with the spectral weight very strongly suppressed, but not exactly vanishing. 
    The insulating gap forms upon decreasing temperature such that at $T=145$ K the system is insulating. 
    The moderate value of $U_{\text{iso}}$ is actually very close to the cRPA value, $U_{\mathrm{cRPA}}$, and well inside the estimated error margin due to the known overestimation of screening processes within cRPA~\cite{cRPA2}.
    Most importantly, at $\Delta U=0.2$ eV, we observe a gap at low temperature, which is in agreement with experiments \cite{Moser, Uchida_ARPES, Optical_conducivity_Ba2IrO4}. 
    Given the experimental evidence of the insulating state in \bairo, we take $\Delta U$ = 0.2 eV to compare our calculations in the next section with experiments.

\section{The electronic structure of \texorpdfstring{$\mathbf{Ba_2IrO_4}$}{Ba2IrO4} }
\label{section:comp_with_exp}

    In this Section we compare the electronic properties of Ba$_2$IrO$_4$ evaluated within DMFT with existing literature, focusing in particular on the spectral function.
    Based on the conclusions of the previous sections, the calculations here are performed on the three-band model at the inverse temperature $\beta=80$ eV$^{-1}$ ($T=145$ K) with a SOC constant $\lambda=0.31 $ eV, and the Coulomb parameters $U$ defined in \autoref{tab:Coulomb_Ba2IrO4} increased by an isotropic amount $\Delta U=0.2$ eV. 
    Computational details of the DMFT calculations and analytic continuations can be found in Appendix~\ref{appendix:Computational_details}. \\
    As already mentioned in \autoref{section:spin_orbit}, the MIT occurs there as a selective Mott transition with filled $j_{\mathrm{eff}}=3/2$ bands and half-filled  $j_{\mathrm{eff}}=1/2$ states, which have well-defined lower and upper Hubbard bands (LHB, UHB).\\
    The DOS of the DFT and DMFT calculations are shown in \autoref{fig:bands_and_cuts}~(a-c). 
    Overall, the DMFT bandwidth is larger than the DFT reference. 
    The additional correlations within DMFT open the Mott gap, leading to vanishing spectral weight at the Fermi energy. 
    Panel (b) reports the different states in the $j_{\mathrm{eff}}$ basis, revealing the opening of the Mott gap in the \jeff\ band. 
    The distance between the maxima of the DOS of the \jeff\ UHB and the LHB (\JEFF\ bands) is about $1.3$ eV ($1.9$ eV). 
    Optical conductivity measurements showed two pronounced peaks in the absorption coefficient, $\alpha$ and $\beta$, which were interpreted as stemming from excitations from the \jeff\ LHB and the \JEFF\ bands to the UHB respectively \cite{Optical_conducivity_Ba2IrO4}.
    Comparing these peak positions to our DMFT results, we find that their distance within DMFT is $\sim0.5\ \text{eV}$ too large with respect to experiments.
    As we will see below, the overestimation of the energy gap can be traced back to an overestimation of the binding energy at the $X$-point.
    Studying the composition of the $j_{\mathrm{eff}}=1/2$ UHB, the contribution of the $d_{xy}$ orbital is found to be lesser than the one of the $d_{xz},d_{yz}$ orbitals, see panel (c). In particular, we have $n_{xz}=n_{yz}=0.807$ and $n_{xy}=0.887$. 
   This anisotropy is consistent with the DFT and cRPA findings discussed in \autoref{subsection:parametrization of the model}.
   For more details on the $t_{2g}$-projected spectral function, see Appendix~\ref{sec:App_t2g_specs}.

    \autoref{fig:bands_and_cuts} (d) shows the total $\mathbf{k}$-resolved spectral function, compared to bands 
    extracted from ARPES measurements in Ref.~\onlinecite{Uchida_ARPES}. 
    We focus first on our DMFT results. 
    Consistently with panel (a), the flat bands around the Fermi level in \autoref{fig:5_vs_3_akw} at $\Delta U=0\ \text{eV}$ have been split into Hubbard bands, opening an indirect band gap between $\Gamma$ and $M$. 
    The UHB exhibits a clear and coherent dispersion over the full $\mathbf{k}$-path. 
    With respect to the DFT Kohn-Sham band structure, the \JEFF\ bands show a negligible renormalization, resulting from a nearly flat imaginary self energy at low frequencies.
    The LHB of the $j_{\mathrm{eff}}=1/2$ state shows instead a rather incoherent behavior, see panel (d). \\

    Comparing with experimental data of Refs.~\onlinecite{Uchida_ARPES,Moser}, we find a nearly perfect agreement around the $M$ point, where the bands have a prominent \JEFF\ character. 
    The corresponding binding energy of $0.35$ eV matches well with experiment ($0.4$ eV \cite{Moser} and $0.26\ \text{eV}$ \cite{Uchida_ARPES}). 
    The good agreement holds for the full dispersion of the \JEFF\ band. 
    The situation is different for the \jeff\ band. 
    In the experimental spectrum, the \jeff\ band has a peak at $X$ with a binding energy in between 0.21 eV \cite{Moser} and $\sim0.3$ eV \cite{Uchida_ARPES}, leading to the gap opening at the $X$ point. 
    In our DMFT simulations, however, the binding energy at that point corresponds to $0.71$ eV. 
    We attribute this mismatch to the absence of antiferromagnetic fluctuations and local moment formation:
    A cooperation of band folding due to the doubling of the unit cell and modifications of the band structure due to the inclusion of short-range correlations should suppress the $j_{\mathrm{eff}}=1/2$ peak at $M$, and shift the peak at $X$ to smaller binding energy. 
    Including non-local correlations in a cluster-DMFT treatment of the paramagnetic phase might improve the agreement with experiment as it was for instance shown for Sr$_2$IrO$_4$ \cite{PhysRevMaterials.2.032001,Lenz_2019}. 
    On the other hand, the completely filled \JEFF\ bands are rather insensitive to these effects.

\begin{figure}[tb!]
\centering
\includegraphics[width=.95\linewidth]{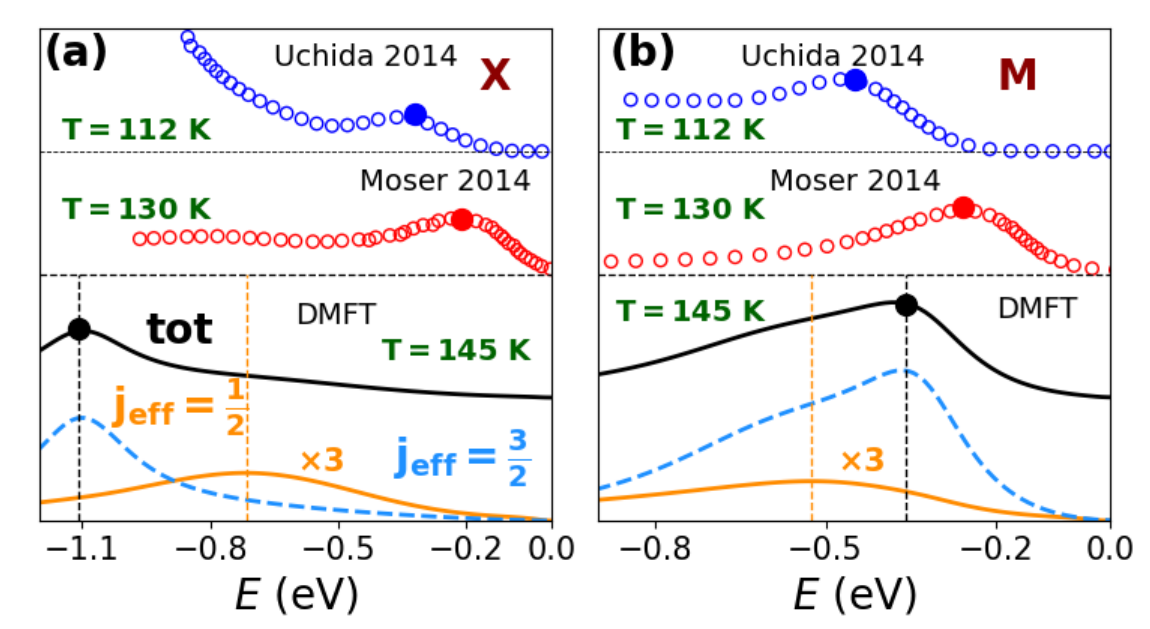}
\caption{
Details of the spectral function.
Electron distribution curves (EDC) of ARPES measurements from Refs.~\onlinecite{Uchida_ARPES,Moser} are compared to the DMFT spectral function at the high-symmetry $\mathbf{k}$-points $X$ (a) and $M$ (b). 
Dots mark maxima of the intensity; in case of the DMFT spectra, contributions of the $j_{\mathrm{eff}}$ states are shown separately. 
}
\label{fig:Akw_zoom}
\end{figure}

    Selected \textbf{k}-resolved spectra at the $X$ and $M$ points are shown in \autoref{fig:Akw_zoom}, where we also plot ARPES data from Ref.~\onlinecite{Moser}. 
    Compared to the thin film samples measured in Ref.~\onlinecite{Uchida_ARPES}, the band structure of bulk samples from Moser \textit{et al.}\cite{Moser} is shifted to larger binding energies, and the energy difference of the features at $X$ and $M$ is smaller. 
    Our difference to experiment consists mainly in two aspects: 
    (i) We overestimate the binding energy at $X$. 
    (ii) The corresponding intensity of the two peaks is quite different, due to the broadening of the $j_{\mathrm{eff}}=1/2$ state. 
    In addition to the total spectral function of our DMFT calculation, we also show its decomposition into $j_{\mathrm{eff}}$ states.
    Due to the incoherent nature of the $j_{\mathrm{eff}}=1/2$ LHB, and despite the binding energy at $X$ having a prominent $j_{\mathrm{eff}}=1/2$ character, the intensity coming from the $j_{\mathrm{eff}}=3/2$ bands is dominating the total spectral intensity. 
    Both peaks at $X$ and $M$ are thus essentially of $j_{\mathrm{eff}}=3/2$ character.  
    For additional plots of the spectral function that allow for a direct comparison with Ref.~\onlinecite{Moser}, we refer the reader to Appendix~\ref{Sec:App_add_specs}.
    \\
\begin{figure}[t!]
\centering
\includegraphics[width=0.95\linewidth]{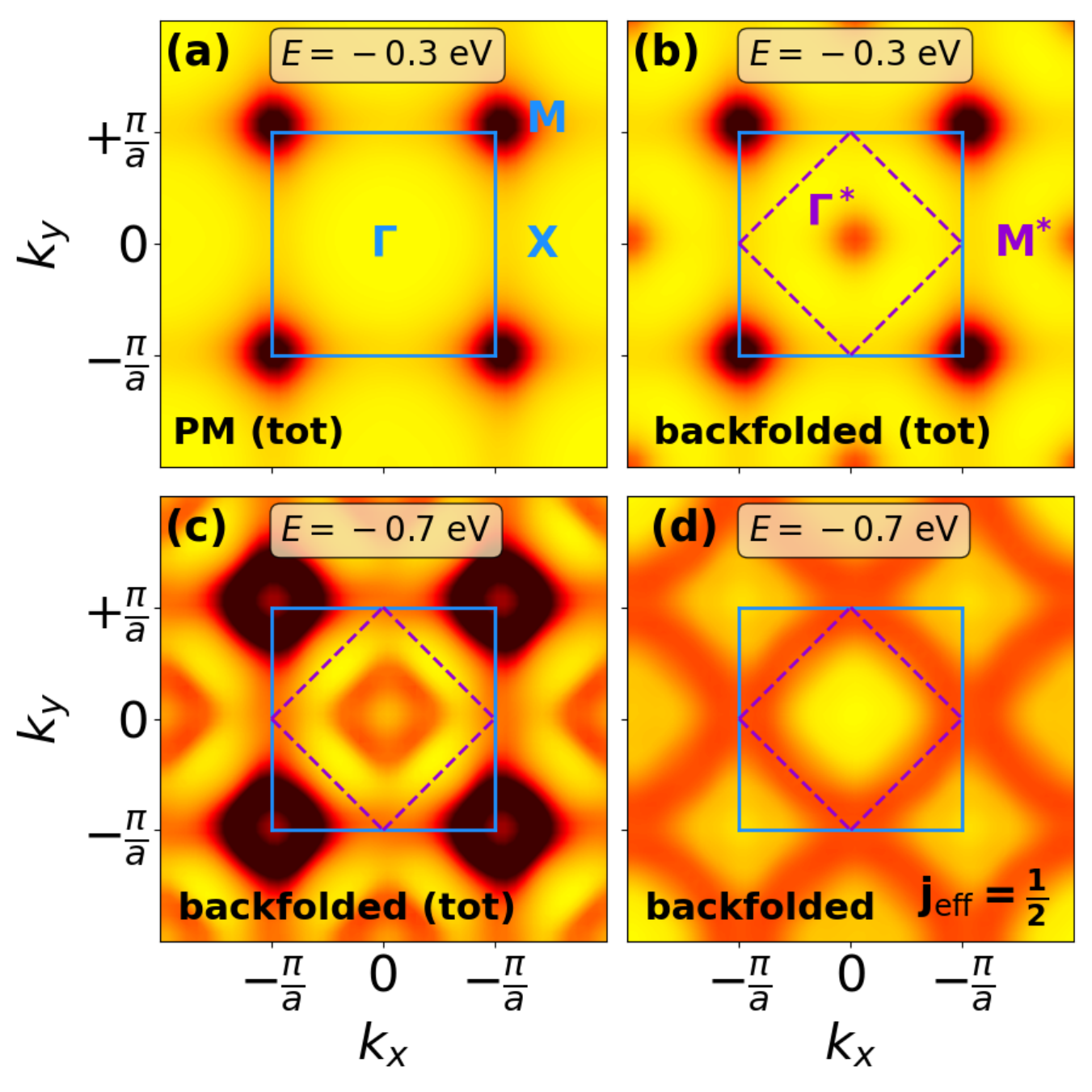}
\caption{ Iso-energy maps of the spectral function.
(a) Color plot of $A(\mathbf{k},E=-0.3 \textrm{ eV})$ in the large BZ indicated in blue. 
In violet, we indicate the small BZ relevant for the antiferromagnetic phase. 
(b-c) CE map with backfolding into the small BZ at $E=-0.3$ eV and $-0.7$ eV respectively. 
(d) Focus on the $j_{\mathrm{eff}}=\frac{1}{2}$ projection for $E=-0.7$ eV. 
To mimic an emerging antiferromagnetic order, panels (b-d) show the spectral weight of $25\%$ of the folded bands and $75\%$ of the unfolded ones.}
\label{fig:constant_energy}
\end{figure}    

    To compare our calculations in more detail with experiment, we show some constant energy (CE) maps in \autoref{fig:constant_energy}.
    For an energy of $-0.3\ \text{eV}$ we only observe large spectral weigth at $M$, consistently with \autoref{fig:bands_and_cuts}. 
    In analogy with the previous discussion, this feature is consistent with Ref.~\cite{Uchida_ARPES}. 
    However, the peak observed in ARPES spectra at $X$ is absent in our calculations. 
    In Ref.~\onlinecite{Moser} a backfolding of the bands is reported with less intensity in the antiferromagnetic BZ. 
    To compare with the spectra in the antiferromagnetic phase, we have to define two different BZs depending on the magnetic state, see \autoref{fig:constant_energy}. 
    There, the different high-symmetry \textbf{k}-points are indicated as $\Gamma^*=\{\Gamma,M\}$, $M^*=X$ in the corresponding colors as well. 
    Even though our DMFT calculations are paramagnetic, we now mimick the effect of antiferromagnetism via backfolding of the bands into the small BZ.
    In order to reproduce an emerging antiferromagnetic ordering, we show in panels (b) and (c) spectra with $25\%$ spectral weight of backfolded and $75\%$ original bands.
    With respect to ARPES spectra in Ref.~\onlinecite{Moser}, our calculation still misses the peak at $X$ ($\mathbf{M^*}$) for low-energy states, but increasing the binding-energy of the energy cuts leads to a better agreement. 
    In the total spectral function, however, the features stemming from the \jeff\ band are heavily suppressed.
    This is due to the large incoherence of the \jeff\ LHB, causing a large difference in intensity with respect to contributions from the \JEFF\ bands.
    We therefore show in panel (d) the projection on the $j_{\mathrm{eff}}=\frac{1}{2}$ band separately. 
    Substantiating the previous discussion, spectral weight is visible in a square around $X$ for the CE map at $E=-0.7$ eV.

\section{Conclusions}
\label{section:summary}
In this work, we simulated the electronic structure of \bairo\ by performing realistic dynamical mean-field theory (DMFT) calculations based on models derived from \textit{ab initio} density functional theory (DFT) and constrained random phase approximation (cRPA) simulations.  
In particular, we presented two models of \bairo\ including Ir-$t_{2g}$ and Ir-$5d$ Wannier functions respectively.
We solved both models within the DFT+DMFT scheme, using a full Coulomb tensor $U$ evaluated from cRPA.
To the best of our knowledge, the five-band Ir-5d calculation is the first presented in literature for compounds with strong spin-orbit coupling (SOC).
Within this framework, we found a good agreement between the two DMFT solutions, validating the use of the simpler $t_{2g}$ model that fully retains the low-energy physics of the system.
Limitations of 
the Ir-$t_{2g}$ model must be searched in the empty part of the spectrum, populated by the upper \jeff\ Hubbard band only.
According to our five-band model, the Ir-$\tilde{d}_{x^2-y^2}$ band is situated close in energy, which limits the predictive power of the Ir-$t_{2g}$ model when it comes to the description of spectroscopies that probe the hole sector of the spectrum.
This might be the case for optical conductivity measurements at energies in the $\sim\text{eV}$ range \cite{Optical_conducivity_Ba2IrO4} , time-resolved ARPES experiments \cite{trARPES} or for probing inelastic excitations, for instance via resonant inelastic x-ray spectroscopy \cite{RIXS-Ba2IrO4}.\\
Using the Ir-$t_{2g}$ model, we studied the interplay between SOC, Coulomb interaction and Hund's exchange, showing how SOC influences the physics of compounds with a $5d^5$ configuration and the ligand-field of \bairo. 
To that aim, we varied the monopole term of the \textit{ab initio} Coulomb tensor $U$, the SOC strength $\lambda$ and the Hund's exchange $J$. 
The resulting phase diagram is remarkably rich.
Three different regions were identified: 
In the weak SOC regime, for $0\leq\lambda<0.1$ eV, the $t_{2g}$ physics dominates and small variations of $\lambda$ imply large changes of the critical interaction strength of the metal-insulator transition, $U_c$.
In contrast, in the large SOC regime, i.e. for $\lambda>0.6$ eV, an effective single-band $j_{\mathrm{eff}}=1/2$ model is realized even with no interactions, and
the critical interaction strength $U_c(\lambda)$ for the Mott transition stays nearly constant. 
The Mott transition is thereby $U$-driven.
At intermediate SOC, i.e. $0.1\ \text{eV}\leq\lambda\leq0.6\ \text{eV}$, a finite value of $U$ with the resulting Hartree shifts is needed to yield an effective single-band \jeff\ picture.
Furthermore, $\lambda$ plays an important topological role: 
We identified two distinct regions in the metallic regime, characterized by a SOC-induced topological (Lifshitz) transition.
Overall, the transition towards the insulating phase can therefore be referred to as spin-orbit Mott transition.

\bairo\ is located in the intermediate SOC region with $\lambda=0.31$ eV. 
This proximity to both the metal-insulator transition and phases with different band topology is intriguing and could guide material design in the future.
Even though in practice, changing the spin-orbit coupling constant $\lambda$  via chemical substitution goes in hand with modifications of the electronic structure and Coulomb interaction as well, we note that reducing $\lambda$ would drive the system to such an interesting point of phase competition.
This could be in principle realised by substituting Ir by $4d$ transition metal atoms like Ru  ($\lambda_{\text{Ru}}\approx0.1\ \text{eV}$ \cite{Tamai2019}).
In Sr$_2$IrO$_4$, however, it is still debated whether a Ru substitution can be interpreted via a reduction in the effective SOC of the material \cite{Zwartsenberg2020,Veronique2021}. \\

At low temperature, we find an insulating state matching with experiment for $U=U_{\mathrm{cRPA}}+0.2$ eV. 
Investigating the Mott transition of \bairo\ at higher temperature, we obtain a crossover resembling the one above the second-order end-point of the Mott transition in the single-band Hubbard model.
We expect the corresponding first order Mott transition to appear for $T\lesssim 150\ \text{K}$. 
Nevertheless, a treatment within a three-orbital framework is essential to properly describe the spectroscopic properties of the system.
This is rooted in the non-negligible orbital mixing within the tight-binding model and the proximity of \JEFF\ bands to the Fermi level, consistent with the strong inter-twining of the \jeff\ and \JEFF\ manifolds observed in optical conductivity measurements \cite{Optical_conducivity_Ba2IrO4}. 
Similar to recent studies for Sr$_2$IrO$_4$ \cite{magnetization_sr2iro4_dxy,Pavarini_jeff_12}, this in particular casts doubts on the validity of the single-band \jeff\ picture upon doping.
Coming back to the parallel of \bairo\ with cuprates, we note that the general question on the validity of an effective one-band model upon doping has been intensely discussed for these systems, i.a. in context of the Zaanen-Sawatzky-Allen diagram \cite{ZSA} and the Zhang-Rice band \cite{ZR1,ZR2,Macridin2005}.
Even though these models rather address the question of the hybridization with oxygen orbitals and the resulting charge-transfer, the hierarchy of different \textit{ab initio} multi-band models with respect to the energy scales that can be correctly captured is similar \cite{Andersen, Hirayama1, Hirayama2}.\\

Comparing the spectral function with existing ARPES experiments, our DMFT solution shows good agreement within the \JEFF\ bands. 
On the other hand, our results do not match perfectly the experimental characterization of the \jeff\ band.
Especially at the $X$ point of the BZ the calculated binding energy is overestimated by $\sim0.5\ \text{eV}$. 
We interpret this difference as a consequence of the absence of non-local correlations and antiferromagnetic fluctuations, which could be included, for instance, via cluster extensions of DMFT. 
We also envision the effective three- and five-band models developed in this study to be important for investigating the change of the \bairo\ spectral function upon doping.

\section{Acknowledgments}

We acknowledge fruitful discussions with Silke Biermann which motivated this work.
We are also grateful to Simon Moser for discussions about ARPES measurements of \bairo, and to Benjamin Bacq-Labreuil for discussions related to DMFT and analytic continuation.
The authors would also like to thank Sophie Beck for her help with using the TRIQS package for DMFT calculations with spin-orbit coupling and Kazuma Nakamura and his team for their help with the cRPA calculation using their code RESPACK. 
This study has been partially supported through the EUR Grant NanoX No. ANR-17-EURE-0009 in the framework of the “Program des Investissements d’Avenir''. 
Part of the calculations for this article was made possible by the access to the HPC resources of CALMIP supercomputing center under the allocation P21048. 
We also acknowledge computation time from the Grand Equipement National de Calcul Intensif provided at IDRIS and TGCC (Project No. A0130912043) and part of the simulations were performed on the SACADO MeSU platform at Sorbonne Universit\'e.
Support from the GDR MEETICC in form of a student mobility grant (F.C.) is gratefully acknowledged.

\setcounter{equation}{0}%
\setcounter{figure}{0}%
\setcounter{table}{0}%
\makeatletter%
\renewcommand{\theequation}{A\arabic{equation}}%
\renewcommand{\thefigure}{A\arabic{figure}}%
\renewcommand{\thetable}{A\Roman{table}}%
\appendix

\section{Computational details}
    \label{appendix:Computational_details}

    The DFT calculations were performed with QuantumEspresso v6.8 \cite{QuantumEspresso1, QuantumEspresso2} using a $8\times8\times8$ Monkhorst-Pack $\mathbf{k}$-point grid centered at $\Gamma$  and a $90\ \text{Ry}$ energy cutoff for the wavefunction expansion with the PBE functional \cite{PBE1996} for Ba$_2$IrO$_4$. 
    We used Optimized Norm Conserving Vanderbilt (ONCVPSP) pseudopotentials \cite{ONCVPSP2013} from the PseudoDojo database \cite{PseudoDojo2018} with (fully relativistic) scalar relativistic corrections for the (non) spin-orbit calculations.
    
    The cRPA calculations were performed using RESPACK v20200103 \cite{Nakamura2021}. 
    The Wannier functions were obtained for the $t_{2g}$ subspace with a $0.8$ coefficient on the initial guess gaussians. 
    The disentanglement was constrained within the $t_{2g}$ bands energy range and a frozen window spanning from $10.40\ \text{eV}$ to $12.55\ \text{eV}$ was chosen. 
    The polarization function was computed using a cutoff of $3.6\ \text{Ry}$ and $100$ bands. 
    The imaginary broadening of the Green's functions was set to $0.1\ \text{eV}$ for the cRPA calculations.

    The DMFT calculations were performed using the TRIQS software \cite{TRIQS2015}, in particular the DFTTools package \cite{TRIQSDFTTOOLS2016} for the $\mathbf{k}$-point summations and the CTHYB impurity solver \cite{TRIQSCTHYB2016}. The calculations of the five-band model in \autoref{section:validation} were converged using $28 \times 10^{6}$ measurements  and $86\times 10^{6}$ measurements were performed
    at the last iterations. For the three-band model in Sec. \ref{section:validation}, \ref{section:spin_orbit} and \ref{section:comp_with_exp} we sampled with $72 \times 10^{6}$ counts over the DMFT loop and we converged the last iterations with $96 \times 10^{6}$ measurements.
    The $\mathbf{k}$-point summations were performed using the Wannier Hamiltonian on the same $8\times8\times8$ Monkhorst-Pack grid as the DFT calculations. 
    In our study we investigated a temperature range from $T=145\ \text{K}$ to $290\ \text{K}$. 
    For the insulating phase, especially within the transition region, convergence has been achieved for $30-40$ iterations.

    The analytic continuations were performed using the Maximum Quantum Entropy Method \cite{MQEM2018} with a smearing factor between $5\times 10^{-3}\ \text{eV}$ and $1\times 10^{-2}\ \text{eV}$.

\section{Optimized atomic coordinates}
    \label{appendix:atom_coordinates}
    In the following, we list the atomic positions within the experimental primitive unit cell, which were optimized using the PBE functional, see \autoref{tab:Ba2IrO4_unitcell}.

    \begin{table}[ht]
        \centering
        \begin{tabular}{c c c c}
        \hline
        \hline
        Unit cell vectors
        & x & y & z \\
        v$_1$ & -2.015 & 2.015 & 6.667 \\ 
        v$_2$ & 2.015 & -2.015 & 6.667 \\
        v$_3$ & 2.015 & 2.015 & -6.667 \\
        \hline
        \hline
        Atom & v$_1$ & v$_2$ & v$_3$ \\ 
        Ba & 0.6456 & 0.6456 & 0.0000 \\
        Ba & 0.3544 & 0.3544 & 0.0000 \\
        Ir & 0.0000 & 0.0000 & 0.0000 \\
        O  & 0.1553 & 0.1553 & 0.0000 \\
        O  & 0.8447 & 0.8447 & 0.0000 \\
        O  & 0.5000 & 0.0000 & 0.5000 \\
        O  & 0.0000 & 0.5000 & 0.5000 \\
        \hline 
        \hline
        \end{tabular}
        \caption{Crystal structure of \bairo.
        Unit cell vectors (in \AA) from Ref.~\onlinecite{Okabe2011} and optimized atomic positions in reduced coordinates.}
        \label{tab:Ba2IrO4_unitcell}
    \end{table}

\section{Interpolation details for the phase diagrams}
\label{appendix:interpolation_phase_diagrams}
In \autoref{section:spin_orbit} of the manuscript we used an interpolation scheme to obtain the two phase diagrams shown.
   To that aim, we started from $A_0(\lambda,\Delta U)$ on a finite grid $(\lambda, \Delta U$), which contains points at the edges of the diagrams and across the transition line. 
   Then, for each $\lambda$ we fitted the profile of $A$ as: 
   \begin{equation}
       A_0(\lambda,\Delta U)=\frac{a(\bar{\lambda})}{1+e^{b(\bar{\lambda})(\Delta U - \Delta U_m(\bar{\lambda}))}}
       \label{eq:fit_G_b_2}
   \end{equation}

   Then, by fitting the parameter $a(\lambda)$,$b(\lambda)$ and $\Delta U_m(\lambda)$ on a denser grid of $\lambda$, we reconstructed the corresponding Fermi function for each value of $a$,$b$ and $\Delta U_m$ on the denser grid.
   We stress that by construction $\Delta U_m$  is exactly the inflection point of the Fermi function.
    Critical iso-value lines are found for a given threshold value by false position method. 
         
\section{DMFT spectral function projected on \texorpdfstring{$t_{2g}$}{t2g} states}
\label{sec:App_t2g_specs}

    \begin{figure}
        \centering
        \includegraphics[width=0.99\linewidth]{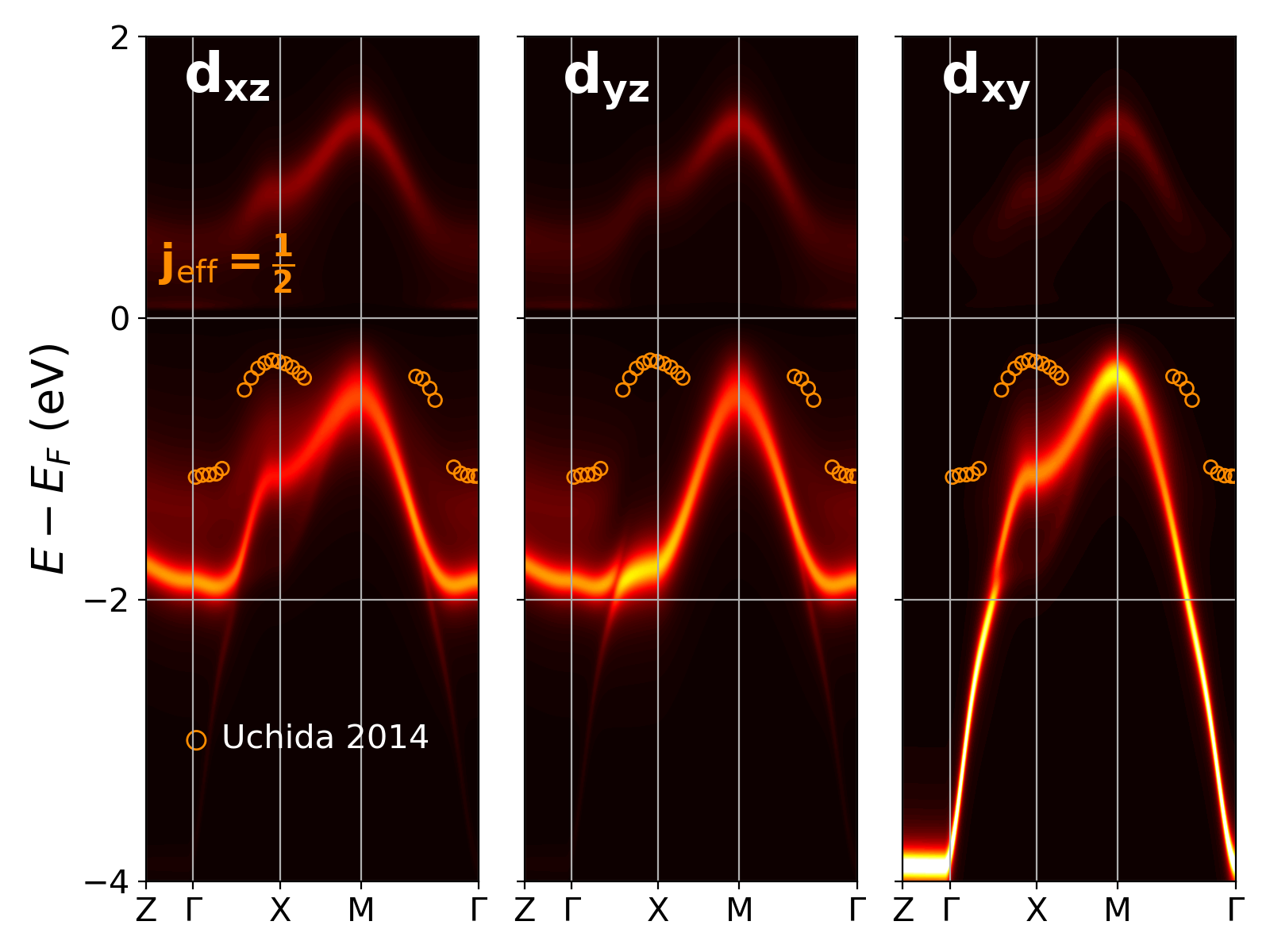}
        \caption{Spectral function in the $t_{2g}$ basis.
        We plot the \textbf{k}-resolved spectral function shown in Fig.~\ref{fig:bands_and_cuts} of the main text, projected onto the $t_{2g}$ orbitals.
        Experimental data points from ARPES measurements of Ref.~\onlinecite{Uchida_ARPES} corresponding to the \jeff\ contribution to the spectrum are shown by orange dots.
        }
        \label{fig:A_w_k_t2g}
    \end{figure}

\autoref{fig:A_w_k_t2g} displays the \textbf{k}-resolved spectral function for $\Delta U=0.2$ eV  projected onto the $t_{2g}$ basis.
To highlight the contribution of the orbital basis with respect to the $j_{\mathrm{eff}} =1/2$ band, we also report data extracted from Ref.~\cite{Uchida_ARPES}. 
We note strong hybridization between the $t_{2g}$ states, especially between $d_{xz}$ and $d_{yz}$, which is expected from SOC, see \autoref{eq:somatrix}. 
Here, we can see that at X (Y) the main contribution to the \jeff\ state comes from the the $d_{xz}$ ($d_{yz}$) orbital. 
In analogy with \autoref{fig:jeffCharacter}, the contribution from the $d_{xz}$ ($d_{yz}$) orbital is only relevant in the path X-M-X (Y-$\Gamma$-Y) and vanishes otherwise. 
\section{Additional plots of the \texorpdfstring{$\mathbf{k}$}{k}-resolved spectral function \texorpdfstring{$A(k,\omega)$}{A(k,w)}}
\label{Sec:App_add_specs}

In this appendix, we report the \textbf{k}-resolved spectral function along additional high-symmetry paths of the BZ, in analogy to the ones reported in  Ref.~\onlinecite{Moser}. 
We performed backfolding to simulate the effect of antiferromagnetic order and mimicked an emerging magnetic order by mixing the intensities of unfolded and backfolded spectra in analogy with  \autoref{section:comp_with_exp}. 
    \begin{figure}
        \centering
        \includegraphics[width=0.99\linewidth]{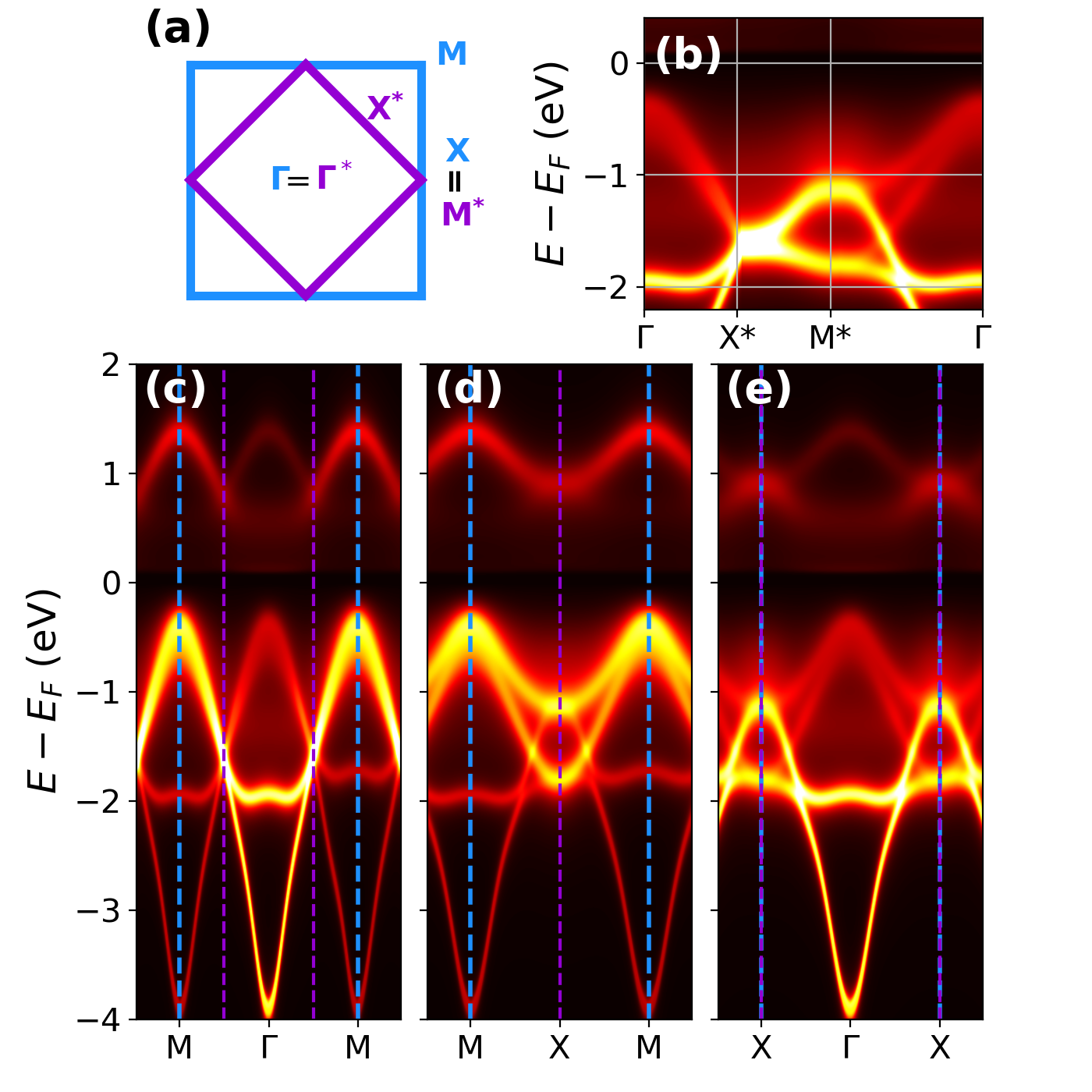}
        \caption{Additional plots of the spectral function of Ba$_2$IrO$_4$.
        (a) Sketch of the different BZ for the paramagnetic (blue) and antiferromagnetic structure (violet).
        (b-e) \textbf{k}-resolved spectral function along different high-symmetry paths, allowing a direct comparison with the spectra published in Ref.~\onlinecite{Moser}.
        }
        \label{fig:comparison_Moser}
    \end{figure}

Panel (a) of \autoref{fig:comparison_Moser} shows the two BZ corresponding to the paramagnetic BZ (blue line) antiferrmomagnetic BZ (in violet). 
We folded the bands in the smaller BZ and mixed $25\%$ of the backfolded bands to $75\%$ of the unfolded ones in analogy with \autoref{section:comp_with_exp}.
Comparing directly to the experimental data of Ref.~\onlinecite{Moser}, we obtain very good agreement concerning the \JEFF\ bands, but the peak at X (M$^*$) stemming from the \jeff\ band is both underestimated in spectral weight and shifted to too high binding energy. 
As discussed in the main text, we attribute this to the absence of antiferromagnetic fluctuations in our DMFT calculations.

 \clearpage
%

\end{document}